\documentclass[%
 reprint,
 amsmath,amssymb,
 aps,
floatfix,
]{revtex4-2}

\usepackage{graphicx}% Include figure files
\usepackage{dcolumn}% Align table columns on decimal point
\usepackage{bm}% bold math
\usepackage{algorithm}
\usepackage[noend]{algpseudocode}
\usepackage{multirow}
\usepackage[utf8]{inputenc}
\usepackage[english]{babel}
\usepackage{xcolor}

\usepackage{soul}

\begin{document}

\newcommand{\hlt}[1]{{{#1}}}

\newcommand{\hlc}[2][yellow]{{\color{#1} {#2}}}

\newcommand{\hlp}[2][purple]{{\color{#1} {#2}}}

\preprint{APS/123-QED}

\title{A Diffusion-Based Embedding of the Stochastic Simulation Algorithm in Continuous Space}% Force line breaks with \\
%%\thanks{A footnote to the article title}%

\author{Marcus Thomas}
 %%\altaffiliation[ ]{Computational Biology Department, Carnegie Mellon University.}%Lines break automatically or can be forced with \\
\author{Russell Schwartz}%
 \email{russells@andrew.cmu.edu}
\affiliation{%
 Computational Biology Department, Carnegie Mellon University
}%

%%\collaboration{MUSO Collaboration}%\noaffiliation

%%\author{Charlie Author}
%% \homepage{http://www.Second.institution.edu/~Charlie.Author}
%%\affiliation{
%% Second institution and/or address\\
%% This line break forced% with \\
%%}%
%%\affiliation{
%% Third institution, the second for Charlie Author
%%}%
%%\author{Delta Author}
%%\affiliation{%
%% Authors' institution and/or address\\
%% This line break forced with \textbackslash\textbackslash
%%}%

%%\collaboration{CLEO Collaboration}%\noaffiliation

\date{\today}% It is always \today, today,

\begin{abstract}%
A variety of simulation methodologies have been used for modeling reaction-diffusion dynamics --- including approaches based on Differential Equations (DE), the Stochastic Simulation Algorithm (SSA), Brownian Dynamics (BD), Green’s Function Reaction Dynamics (GFRD), and variations thereon --- each offering tradeoffs with respect to the ranges of phenomena they can model, their computational tractability, and the difficulty of fitting them to experimental measurements.  Here, we develop a multiscale approach combining efficient SSA-like sampling suitable for well-mixed systems with aspects of the slower but space-aware GFRD model, assuming as with GFRD that reactions occur in a spatially heterogeneous environment that must be explicitly modeled. Our method extends the SSA approach in two major ways. First, we sample bimolecular association reactions following diffusive motion with a time-dependent reaction propensity. Second, reaction locations are sampled from within overlapping diffusion spheres describing the spatial probability densities of individual reactants. We show the approach to provide efficient simulation of spatially heterogeneous biochemistry in comparison to alternative methods via application to a Michaelis-Menten model. 

\end{abstract}

\maketitle

%\tableofcontents

\section{INTRODUCTION}
 Simulation methods have become a valuable adjunct to experimental work, facilitating the interpretation of experimental data and inferences about experimentally unobservable aspects of biomolecular dynamics \cite{kaya2018heterogeneities}, yet accurate simulations remain challenging for many biochemical processes crucial to living systems.   The need for improvements in simulation technology is particularly acute for macromolecular assembly systems, which are central to nearly all cellular processes, yet frequently not directly observable experimentally due to their small scale and rapid dynamics \cite{thomas2017quantitative}. Intractability of experimental approaches is particularly acute for understanding self-assembly {\em in vivo}, which may operate quite differently from purified {\em in vitro} models due to such effects as spatial confinement \cite{chevreuil2018nonequilibrium,lopez2018ship,wang2018interplay}, macromolecular crowding \cite{junker2019impact, smith2014applying}, and influences of extrinsic cellular factors \cite{van2018rna}.
 %Cellular signalling and regulatory networks present similar difficulties in terms of their spatiotemporal characterization with purely experimental investigation, as large numbers of possible postranslational modification states and complex spatial dynamics make such systems challenging to reproduce with simplified mathematical models. [NEED A CITATION IF THIS SENTENCE IS INCLUDED]
 The challenges of developing simulations that are both accurate and efficient, especially for hard-to-model systems like self-assembly, has led to extensive work on models and algorithms for biochemical simulation seeking to balance computational efficiency with fidelity to the complexity of the underlying biology.

The Gillespie Stochastic Simulation Algorithm (SSA) \cite{gillespie1992rigorous,gillespie2009diffusional} was particularly influential in establishing a computational framework for efficient sampling of chemical reaction trajectories, especially for small copy-number settings typical of biochemistry in the cell.  The SSA has proven a valuable tool for understanding the kinetics of reaction networks, i.e., tracking the evolving populations of interacting reactant species, when older methods based on deterministic differential equation systems are too inaccurate or computationally infeasible \cite{gillespie2013perspective,warne2019simulation,nag2009aggregation}.  Many improvements have been made to efficiency of the basic method either via approximations or for particular spaces of model parameter \cite{gillespie2001approximate,rathinam2003stiffness, cao2005slow, jamalyaria2005queue,misra2008efficient,  anderson2008incorporating,sneddon2011efficient,donovan2013efficient,lin2019scaling}.  Yet the SSA is not explicitly spatial and instead treats the reactants as uniformly distributed at all times, aside from transient fluctuations. To better capture spatial heterogeneity, extensions of the SSA have been developed based on the reaction diffusion master equation (RDME), typically partitioning the reaction volume into compartments or voxels for which the usual well-mixed assumption applies in each compartment \cite{baras1996reaction,isaacson2009reaction}. In these spatial Gillespie models, reactants can react within a compartment or diffuse to an adjacent compartment. However, there is an inherent conflict between accuracy (smaller compartments imply higher spatial resolution) and the well-mixed assumption (better satisfied with larger compartments and/or diffusion rates). In fact, even in the limit of fast diffusion rates, RDME may not converge to the Chemical Master Equation (CME) underlying the Gillespie algorithm \cite{smith2016breakdown}.

Brownian dynamics (BD) methods provided an opposite extreme of efficiency/realism tradeoffs for such modeling, allowing detailed, off-lattice spatial dynamics but at much greater computational cost.  Coarse-grained BD methods have been widely used in self-assembly modeling, as they can deal well with systems with complicated spatial heterogeneity or geometrically intricate structures \cite{schwartz1998local,bourov2003role,hagan2006dynamic,kerr2008fast,castle2013brownian,castro2014brownian,bachmann2016bond,donev2018efficient}. However, their need to explicitly model diffusion trajectories of single particles creates high computational demands due to the large gap between timescales of diffusive motion versus those of typical molecular assembly processes. 
Smoldyn \cite{andrews2010detailed} is one particularly prominent example, in which molecules diffuse with ideal Brownian motion and react upon collisions. Smoldyn has been considerably extended and improved since its initial release in 2003, e.g., by the inclusion of rule-based modeling, volume exclusion handling, on-surface diffusion, single particle tracking, and integration with BioNetGen. However, Smoldyn depends on the use of discrete fixed time steps, creating tradeoffs between accuracy and efficient run time in some problem domains.

Green's function reaction dynamics (GFRD \cite{van2005green,van2005simulating}) provided an alternative approach to capture spatial heterogeneity in simulating reaction-diffusion systems while taking advantage of SSA-like efficient discrete event simulation without requiring spatial discretization. Instead of generating sample trajectories from the CME or RDME through MCMC, or numerically solving the many-body Smoluchowski equation as in Brownian Dynamics, GFRD analytically solves the Smoluchowski equation for single molecules and molecular pairs in terms of Green's functions. These Green's functions describe the probability of finding a molecule (pair) at a certain location and time given a known position(s) at an earlier time. A maximum time step is chosen such that, with high probability, at most two molecules come into contact, a requirement for analytical tractability. This single/pairwise interaction assumption becomes more valid with smaller time steps, introducing a trade-off between accuracy and efficiency. Reactions are incorporated through the boundary conditions, and the method combines into a single step propagation through space and reactions between particles. eGFRD \cite{sokolowski2019egfrd} is a more recent exact algorithm which removes the accuracy/efficiency trade-off by including the concept of "protective domains" first developed by Oppelstrup et al. \cite{opplestrup2006first}. These domains are geometrically simple mathematical boundaries enclosing single molecules or pairs, each of which requires a distinct Green's function solution yielding next event types (domain escape or reaction) and waiting times. Because the time steps are now domain specific, eGFRD is an asynchronous algorithm allowing increased efficiency in some circumstances, although the additional mathematical complexity comes at significant computational expense.

The Small Voxel Tracking Algorithm (SVTA) \cite{gillespie2014small} offers another strategy for particle-based simulation of reaction-diffusion systems.  While SVTA is based on the same underlying physics as eGFRD, its implementation is based on a discrete space model. Instead of protective domains, SVTA constructs one and two particle ``corrals,'' within which single molecules and molecule pairs hop between voxels and potentially interact. More specifically, it is the center of each molecule that hops, since the voxel size is typically smaller than the molecular radius. These small voxel dimensions rule out the use of traditional bimolecular propensity functions that rely on the well-mixed assumption. Because the system state evolves on the time scale of diffusion hops, SVTA does not need to analytically sample locations on the protective domains, an easy task only when the domain is a sphere or other simple shape. It can instead simply keep track of when a diffusion hop places a molecule's center in a voxel identified with the corral. SVTA therefore bypasses the need for a suite of domain specific Green's functions in favor of implementing individual diffusion steps on a lattice, providing a strategy for fast but physically realistic sampling compared to prior off-lattice alternatives.

Similarly, the Microscopic Lattice Method (MLM) \cite{chew2018reaction} simulates lattice-based diffusion with reactions. However, MLM aims at optimizing efficiency by simulating molecules of equal size, and requires that voxel dimensions are larger than molecular radii. Additionally, without corrals or protective domains, MLM relies on periodic boundaries to control the simulation volume and number of molecules. A direct comparison between Chew's MLM and Gillespie's SVTA is unavailable, although each would appear to offer some advantages, the former particularly with respect to efficiency while latter can currently simulate more complex biochemistry. 

Despite these advances, the most challenging systems remain out of reach of molecular simulation methods without substantial simplifications \cite{thomas2017quantitative}.  New advances in models and algorithms for efficient but physically realistic simulation remain a pressing concern if the field is to continue to move towards solving the grand challenge of truly comprehensive and predictive models of whole-cell biochemistry.

In the present work, we develop an alternative methodology intended to reduce the computational complexity of eGFRD while maintaining discrete event based system updates. The method makes use of a Green's function representation of possible particle positions as a function of time, as originally proposed in GFRD \cite{van2005green}, but with an alternative formulation of the probability function in terms of joint probabilities densities of pairs of interacting particles simultaneously.  This reformulation enables a new sampling algorithm for position updates that allows for different tradeoffs of efficiency and precision with current alternatives such as eGFRD, SVTA, and Smoldyn.  Our goal is not to present a fully optimized algorithm or simulation tool, but rather to explore and advance an alternative approach for the use of time-dependent reaction propensities as a basis for reaction-diffusion simulation in continuous space, which may offer better trade-offs between realism and efficiency than prior methods in some problem domains. 

\section*{Theoretical Framework}
\label{Theory}
In this section, we present some theoretical concepts that will be useful subsequently in explaining our model and its relationship to prior work.
Consider the bimolecular association reaction system:
$$ A + B \Rightarrow C $$
governed by
\begin{equation}
\frac{d[A](t)}{dt} = \frac{d[B](t)}{dt} = -k(t)[A](t)[B](t) 
\end{equation}
where A and B are hard-sphere species with radii $r_{A}$ and $r_{B}$ and diffusion coefficients $D_{A}$ and $D_{B}$. 
There are two traditional treatments of diffusion influenced reactions.  The first was introduced by Smoluchowski \cite{von1917mathematical} and later extended by Collins and Kimball (CK) \cite{collins1949diffusion}. At time t=0, a single particle of species A is considered fixed at the origin and an initial surrounding concentration gradient is set up for the mobile species B molecules. They showed that
\begin{equation}
k(t) = \Phi(t)/c_{0} = (4\pi R^{2}D/c_{0})(\partial c/\partial r)_{r=R}
\end{equation}
where $\Phi(t)$ is the probability flux across a boundary sphere for the A particle at radius R, and $c_{0}$ is the initial uniform concentration for species B. The simultaneous diffusion of both species is incorporated by setting D as the sum of their respective diffusion coefficients. In this picture, the concentration gradient for the mobile B species 
%EDIT
$c(r,t)$, defined as the concentration of the B species at distance $r$ from the origin at time $t$ after the initial condition,
%EDIT
is found by solving the diffusion equation
\begin{equation}
\partial c/\partial t = D\nabla^{2}c
\end{equation}
subject to initial condition $c(r,0) = c_{0}$ and the radiation boundary condition 
% EDIT
$D(\partial c/\partial r)_{r=R} = \kappa c(R,t)$ where $\kappa$ is a specific reaction rate.
%EDIT
The solution $c(r,t)$ is a complicated function and obeys the relation
\begin{equation}
k(t)/k_{i} = \frac{c(R,t)}{c_{0}}
\end{equation}
where $k_{i}$ is the limiting value $k(t\Rightarrow 0)$. Naqvi et al.\cite{naqvi1982kinetics} (sections III.-IV.) updates this by replacing the diffusion equation with a discrete random walk model from which is obtained in the limit of sufficiently long time and distance scales
\begin{equation}
k(t)/k_{0} = \frac{c(R+\Delta,t)}{c_{0}}
\end{equation}
with $\Delta$ equal to two thirds the scattering mean free path.

The second treatment is due primarily to Noyes \cite{noyes1956models} and considers an isolated pair of reactive molecules separating from a nonreactive encounter. They showed that

\begin{equation}
k(t) = k_{0}\Big[1 - \int_{0}^{t}  h(t')dt'\Big] 
\end{equation}
where $k_{0}$ is defined as ``the rate constant applicable for an equilibrium molecular distribution''\cite{noyes1961prog} and $h(t)dt$ is the ``probability two molecules separating from a nonreactive encounter at time zero will react with each other between $t$ and $t+dt$'' \cite{noyes1956models}. This can be recast into the form (\cite{naqvi1982kinetics} Eq.47)
\begin{equation}
k(t)/k_{0} = S(t;r_{0}=R_{0},R)
\end{equation}
where $R_{0}$ denotes the distance between two molecules separating from a nonreactive encounter at time zero, and the survival probability $S(t;r_{0},R)$ is defined as
\begin{equation}
S(t;r_{0},R) = 1 - \int_{0}^{t}p(t';r_{0},R)dt'.
\end{equation}

These two major approaches, based on the diffusion equation and particle-pair standpoint respectively, can be shown to be equivalent under certain assumptions and by a lengthy derivation (see \cite{naqvi1982kinetics}, sections IV and V). Our method is most easily identified with the theoretical framework of Noyes, but with a different emphasis on instantiating the physical model so as to enable efficient stochastic off-lattice particle simulations.  We describe the novel features of our model in more detail below.\\  
The function $h(t)$ appearing in Noyes' fundamental relation can be inferred as the special case
\begin{equation}
h(t) = p(t;r_{0}=R_{0},R)
\end{equation}
To be clear, the initial separation $r_{0}$ is the separation distance immediately after a nonreactive encounter. Naqvi argues that $r_{0} \neq R$, the reactive contact distance defined in the boundary condition, but instead $r_{0} = R_{0} = R + \Delta$. The exact expression for $p(t;r_{0},R)$ depends on various assumptions, e.g., that the discrete random walks taken by the particles are accurately described by a continuous diffusion equation. In this case, one needs to make further assumptions about initial conditions and boundary conditions.

In the CK picture, the reaction rate evolves only during the time window beginning with the initial condition and ending with a reaction. The assumption here is that immediately after a reaction, the system returns the concentration surrounding the product molecule to the fixed initial value. As such, the formalism may not be suitable to an event-driven, explicitly spatial simulation. Chew et al. \cite{chew2018reaction} with their microscopic lattice method address this issue by deriving their lattice parameters as analogues to the effective or steady state reaction rates in the continuum CK/Noyes theory. This ensures the model behaves similarly to the theory over suitably long time scales.

While our treatment of diffusion influenced reactions is similar to the particle pair approach in Noyes theory, there are notable differences. Instead of using probabilistic arguments to derive reaction rate functions suitable for a differential equation model, we use them to derive reaction propensities suitable for a discrete event SSA. Our conception is as follows: Given a collection of molecules in an explicit and bounded 3d space, and assuming a maximum diffusion time before which we observe their positions, reaction waiting times can be randomly sampled using pairwise propensity functions. The probability density we focus on is not $h(t) = p(t;r_{0}=R_{0},R)$, but rather $p(t;r_{0},R)$ where the initial separation $r_{0}$ is specified for each molecule pair, and $R$ is the center to center distance at which a reaction can occur.

In the remainder of the paper we describe the model and implementation, which we refer to as the Diffusion-Based Embedding of the Stochastic Simulation Algorithm in Continuous Space (DESSA-CS) method, in reference to an earlier space-free method \cite{zhang2005implementation} based on an accelerated SSA algorithm \cite{jamalyaria2005queue}, and demonstrate its effectiveness in comparison to prior alternatives through application to a Michaelis-Menten model.

\section{Methods}

Algorithm \ref{alg:snrm} summarizes our general procedure for off-lattice spatial stochastic simulation. It makes use of a discrete event structure similar to the stochastic simulation algorithm, with the addition of routines for sampling reaction locations. This sampling is based on diffusion spheres containing $n_{sigma}$ standard deviations of the Gaussian distributions describing each particle, similar to GFRD. The resulting positions (due to reactions and position-only updates) are therefore restricted to be within the diffusion spheres, no matter the choice of $n_{sigma}$ (typically 3-5).

In contrast with existing simulation methods in which the boundaries of the simulation volume are either periodic or reflective, we utilize an alternate approach. The state of each molecule is represented as a probability distribution, therefore we only have access to precise positions immediately following an event, and do not consider velocities at all. For this reason, traditional periodic and reflective boundaries are not well defined. Our approach to reaction location sampling, assuming the waiting time has been accurately sampled previously, is to allow the diffusion spheres of molecules near the boundary to extend some distance beyond the boundary, typically a fraction of the container length. If the sampled location happens to be outside the container, we implement a reflection procedure designed to keep the molecules within the simulation volume while respecting the physics of diffusion.

\begin{algorithm}[H]
    \caption{DESSA-CS procedure}
    \label{alg:snrm}
    \begin{algorithmic}[1]
            %\Procedure{MyProcedure}{}
            \State Initialize Event Queue: For each assembly, consider self events (unimolecular reaction, position-only update) and pair events (bimolecular reaction) and add to the queue the earliest self event and pair event for each assembly.
            \State Main Loop:
            \Repeat
            \State Extract the next event on the queue.
            \If{event is bimolecular and valid} 
            \State{sample location for product given waiting time; update data structures; add next self event(s) to the queue; add next potential bimolecular events to the queue.}  
            \ElsIf{event is unimolecular event and valid}
            \State{sample locations for both products; update data structures; add next self event(s) to the queue; consider each product and add next potential bimolecular events to the queue.}
            \ElsIf{event is position-only update and valid}
            \State{sample location; update data structures; add next position-only update to the queue; add next potential bimolecular events to the queue.}
            \EndIf
            \State (Apply boundary conditions to product(s) if necessary, before adding new events to the queue.)
            \Until{max allowed simulation time or max allowed number of reactions is reached}
            %\EndProcedure
    \end{algorithmic}
\end{algorithm}

\section*{Sampling Bimolecular Reaction Waiting Times}
Consider a set of $K$ possible bimolecular reactions, i.e., distinct pairs of individual molecules represented as either point particles or finite spheres, and assume each molecule traverses an explicit 3d space by diffusion. For each molecule pair, $k$, there exists a reaction propensity $a_{k}(t;s)dt$ describing the probability of an encounter and subsequent reaction of that pair, within some small time interval [$t$, $t+dt$) after the most recently executed event at time $s$. The waiting time, $t_{wait}$, before the next reaction of reactant pair $k$ can be sampled via the equation \cite{anderson2007modified}
\begin{equation} \label{wait_time_sampling_integral}
\int_{0}^{t_{wait}} a_{k}(t \hspace{1mm}|\hspace{1mm} s)dt = ln(1/r_{k}) 
\end{equation}

\noindent
which determines the time at which the integrated propensity equals an exponentially distributed random variable. 
%EDIT
$r_{k}$ is the uniform random number drawn for molecular pair $k$, for use in sampling an exponential waiting time by the transformation method.
%EDIT
Because each of our propensity functions are unique to their associated molecular pair, the reaction channels defined in the original SSA and in Anderson's modified next reaction method \cite{anderson2007modified} at the the species level are now defined at the molecule pair level.

\subsection*{Point Particles}

At the moment a given molecule's state is updated, the probability density describing its center of mass is concentrated at a single point, i.e., a Dirac delta function centered on that point. As time progresses, the probability density spreads as a Gaussian. This is the free diffusion Green's function solution of the diffusion equation \cite{van2005green}. The positions of two molecules A and B are therefore described by two independent Gaussian random variables, $x_{A}(t)\sim\ N[\mu_{A},\Sigma_{A}(t)]$ and $x_{B}(t)\sim\ N[\mu_{B},\Sigma_{B}(t)]$. In order to evaluate Pr($encounter$ \& $reaction$ $|$ $t$), the joint probability of an encounter and a reaction during the interval $[t,t+dt)$, we factor the joint probability as Pr($encounter$ $|$ $t$) * Pr($reaction$ $|$ $encounter$). The latter factor is expressed using a time-independent \textit{intrinsic} reaction rate constant, $c$, such that $c \hspace{1mm} dt$ is the constant encounter conditioned reaction probability over a small time interval. The constant $c$ is specific to this point particle formalism and not equivalent to the microscopic reaction rates used in Smoluchowski or Collins-Kimball theory.

In evaluating the former factor, Pr($encounter$ $|$ $t$), we assume the initial positions of A and B are known and ask the following question: given a sampled position $\textbf{x}_{A}$ of molecule A taken after time $t$, what is the probability a sampled position $\textbf{x}_{B}$ of molecule B after time $t$ will be close to A? Here "close" means at a distance less than a threshold denoting contact or an encounter.

This question can be answered in the language of distributions of quadratic forms in random variables. We define the quadratic form $Q(t)$ as the squared Euclidean distance between Gaussian random variates $\textbf{x}_{A}$ and $\textbf{x}_{B}$.
$$X_{B-A}(t) \sim\ N\big(\mu_{B} - \mu_{A},[\Sigma_{A}(t) + \Sigma_{B}(t)]\big)$$
\begin{equation}
Q(t) = X_{B-A}(t)^{T} \hspace{1mm}  X_{B-A}(t)
\end{equation}
Thus, 
\begin{align} \label{Pr_encounter}
Pr(encounter  \hspace{1mm}|\hspace{1mm} t) &= CDF_{Q(t)}(R_{enc}^{2})\\
&= Pr\big(Q(t) < R_{enc}^{2}\big)
\end{align}
where $R_{enc}^{2}$ is the square of the encounter threshold distance.
Theorem 4.2b.1 of Mathai \& Provost\cite{mathai1992quadratic} provides a formula in terms of an infinite power series expansion which we use for evaluation. 

\begin{equation} \label{CDF_Q_eq}
CDF_{Q(t)}(R_{enc}^{2}) = \sum_{h=0}^{\infty} (-1)^{h} z_{h}(t) \frac{(R_{enc}^{2})^{(3/2)+h}}{\Gamma \big((3/2) + h + 1 \big)}
\end{equation}
The coefficients $z_{h}(t)$ are defined recursively and depend on $\mu_{AB} = \mu_{B} - \mu_{A} $, and $ \Sigma_{AB}(t) = (\Sigma_{A} + \Sigma_{B})$. See Appendix \ref{A1} for a full description. Convergence is defined by no change to 5 places after the decimal for 20 successively higher order approximations. For very small $t$ and large initial separation, the approximation can oscillate wildly about zero. In these parameter regions where numerical instability is detected, we set the CDF to zero. 

\iffalse
Unfortunately, the finite sum approximation does not always converge in a qualitatively similar way as more terms are included, though it is generally more well-behaved with $R_{enc}^{2}<1$. Frequently, it will converge after a few terms ($n_{terms} \sim 10$) and then begin to diverge after a large number of terms ($n_{terms} \sim 100$). However, it may also oscillate wildly with few terms, begin to converge for a while, and then diverge again with oscillatory behavior after many terms are included. We currently define convergence as no change to 5 places after the decimal for $\sim 10$ successive approximations.  However, future work might explore the use of alternate expressions for $CDF_{Q(t)}(R_{enc}^{2})$ that converge with fewer terms, beginning with the formulation developed in Gideon and Gurland\cite{gideon1976series}. 
\fi

With isotropic diffusion, the reaction propensity given $R_{enc}^{2}$ and $d = norm(\mu_{AB})$ after time $t$, and with intrinsic rate $c$, can be reparameterized as a function of the variance $v$ of $X_{B-A}(t)$ rather than of time directly. This variance is simply the diagonal element of $\Sigma_{AB}(t)$.  
%EDIT
The reparameterized reaction propensity, denoted $a_k(t)$, is then given by:
\begin{equation}
a_{k}(t)dt = a_{k}(v \hspace{1mm}|\hspace{1mm} d_{k},R_{enc,k}^{2},c)dv
\end{equation}

\begin{figure*}[!]
  \includegraphics[width=.8\textwidth]{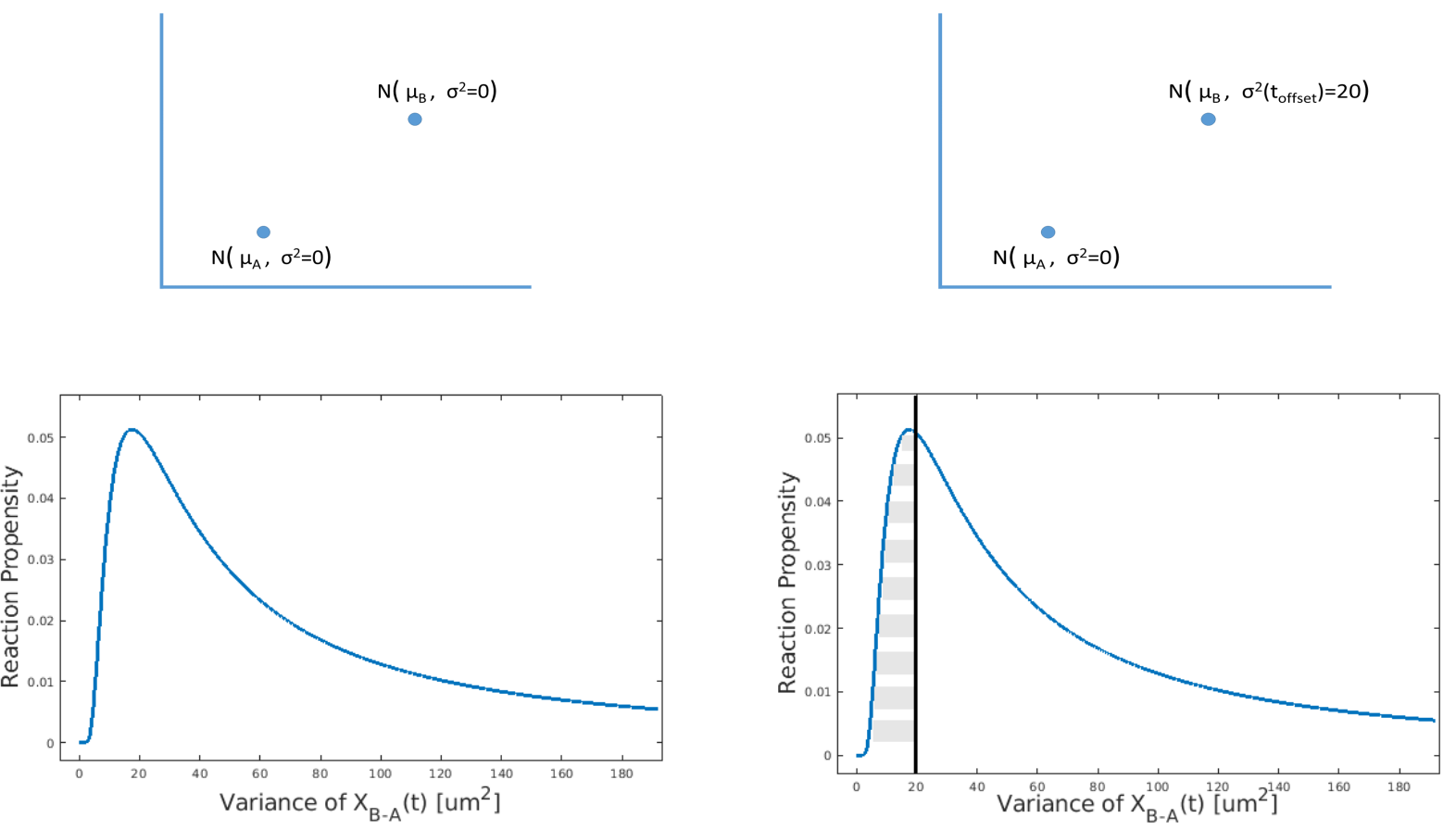}
  \caption{The figures on the left depict two molecules, A and B, described as Gaussians with means separated by $d = 7.235 \mu m$. The point-particle reaction propensity grows as the variance increases, reaching a peak just before $20\mu m^{2}$ and then decreases monotonically. The figures on the right are more typically encountered in the algorithm. The most recent reaction for A was just executed and wait times are being sampled for the $A+B$ reaction. B has already been diffusing for a time $t_{offset}$, thus, propensity function integration begins not at zero variance, but instead at variance equal to $20 \mu m^{2}$.  }
  \label{fig:discounted_integrated_propensity}
\end{figure*}

%EDIT
The time to next reaction can now be determined by evaluating 

\begin{equation} \label{argmin_a}
    argmin_{v} \hspace{3mm} \int a_{k}(v)dv \geq ln(1/r_{k})
\end{equation}

\noindent
and inferring $t_{wait}$ from the variance value, should it exist. Figure \ref{fig:Combo__successful_and_unsuccessful} visualizes the wait time sampling procedure. One added complication is that the DESSA-CS algorithm is event driven. After each event, potential new reactions are considered for the product(s) of that most recently executed event. This implies that the position of the product (e.g., reactant A) is known precisely, while its potential partner (e.g., reactant B) has been diffusing for a time $t_{offset}$ and thus has its position represented by a Gaussian random variable. Any integrated propensity up through $v(t_{offset})$ must therefore be discounted when sampling the variance at which a reaction occurs. See Figure \ref{fig:discounted_integrated_propensity} for an illustration.
The sampling procedure is described in Algorithm \ref{alg:sbwt}. For finite sized molecules, the procedure is similar, except the integrated propensities (called $IntF$ in Algorithm \ref{alg:sbwt}) are expressed directly in terms of times rather than variances. A Matlab implementation of the algorithm is available on GitHub \cite{dessaCS_github}.

%\iffalse
Our propensity function describing Pr($encounter$ \& $reaction$ $|$ $t$) is equivalent to $p(t;r,R)$ from the Noyes theory under the assumption that the molecules are dimensionless point particles for which there is no minimum separation distance. In this case, there is no need to go beyond the free diffusion Green's function solution to the diffusion equation as there are no boundary conditions enforcing a minimum pairwise separation. 

\subsection*{Particles with Finite Size}
With molecules of finite size, we will still use integrated reaction propensities to sample reaction waiting times. However, a different mathematical framework is required to construct the propensities. Assume both particles are spherical and $R$ defines the center-to-center distance at contact. In this context, $p(t;r_{0},R)$ described in the Theoretical Framework section above is expressed as
\begin{equation}
p(t;r_{0},R) = p(r=R,t;r_{0},0) * c \hspace{1mm}dt
\end{equation}
where $c$ now denotes the absorbing/radiation boundary condition parameter, and $p(r,t;r_{0},0)$ is the Green's function solution to the following boundary value problem.  
%The propensity function is again constructed as Pr($encounter$ \& $reaction$ $|$ $t$), however we now express the probability of an encounter given an initial separation as Pr($encounter$) $= p(R,t;r_{0},0)$. 
Assume $p(r,t;r_{0},0)$ obeys a diffusion equation, and the initial separation between molecules is $r_{0}$. This is expressed with the initial condition
\begin{equation}
p(r,0) = \frac{\delta(r-r_{0})}{4\pi r^2}.
\end{equation}
The two boundary conditions on $p(r,t;r_{0},0)$ ensure that the molecular separation never reaches infinity, and that at contact, the probability of a reaction is accounted for:
\begin{align}
    \lim_{r\to \infty} p(r,t) &= 0\\
    4\pi R^{2} D \hspace{1mm} \frac{\partial p(r,t;r_{0},0)}{\partial r} \Biggr|_{r=R} &= c \hspace{1mm} p(R,t;r_{0},0)
\end{align}
From Chew et al. \cite{chew2018reaction}, and Jaeger \& Carslaw \cite{carslaw1959conduction} p. 368), the Green's function solution is
\begin{multline} \label{greensFunction}
    p(r,t;r_{0},0) = \frac{1}{8\pi r r_{0}} \frac{1}{\sqrt{\pi D t}} \Big( exp[-(r-r_{0})^2/4Dt]\\ + exp[-(r+r_{0}-2R)^{2}/4Dt]\\
    - 2B\sqrt{\pi D t} \hspace{1mm} exp[B^{2}Dt + B(r+r_{0}-2R)] \hspace{1mm}\\ *erfc(\frac{(r_{0}-R)}{2\sqrt{Dt}} + B\sqrt{Dt}) \Big)
\end{multline}
where $B = (1+\frac{c}{4\pi R D})/R$. Note that the Green's function also depends on $c$ through $B$.
The propensity function is 
\begin{equation}
a(t)dt = p(R,t;r_{0},0) \hspace{1mm} c \hspace{1mm} dt
\end{equation}
and the time to next reaction, $t_{wait}$, can be determined from the integrated propensity by evaluating
% Wolfram alpha:
%integral  ( (1/sqrt(pi*t))*exp(-(R-r)^2*B^2 / (4*t)) - exp(t)*exp(B(r-R))*erfc(sqrt(t) + (r-R)*B/(2*sqrt(t)) ) )  dt 

\begin{multline} \label{integrated_propensity}
    \frac{B}{4\pi R^2 r_{0}} \hspace{1mm} \Bigg[ erfc\big[ \frac{B(r_{0}-R)}{2\sqrt{\tau}} \big] -\\
    \Big( exp(Br_{0} - BR + \tau)\hspace{1mm} erfc\big[ \frac{Br_{0} - BR + 2\tau}{2\sqrt{\tau}}\big] \Big) - 1 \Bigg]_{0}^{\tau-max}\\
    - ln(1/r_{k}) = 0
\end{multline}

\noindent
with $r_{k} \sim uniform[0,1]$ and $\tau = tDB^{2}$. The waiting time is inferred as $t_{wait} = \tau/DB^{2}$. As in the point particle context, when molecule B of the molecular pair has been diffusing for a time $t_{offset}$ when we are sampling reactions for molecule A, the integrated propensity up through $t_{offset}$ must first be subtracted from the L.H.S. of Eq. \ref{integrated_propensity}. Alternately, Eq. \ref{greensFunction} can be numerically integrated.

\subsection*{Validation of the Wait Time Sampling Procedure}
\begin{figure*}[!]
  \includegraphics[width=.8\textwidth]{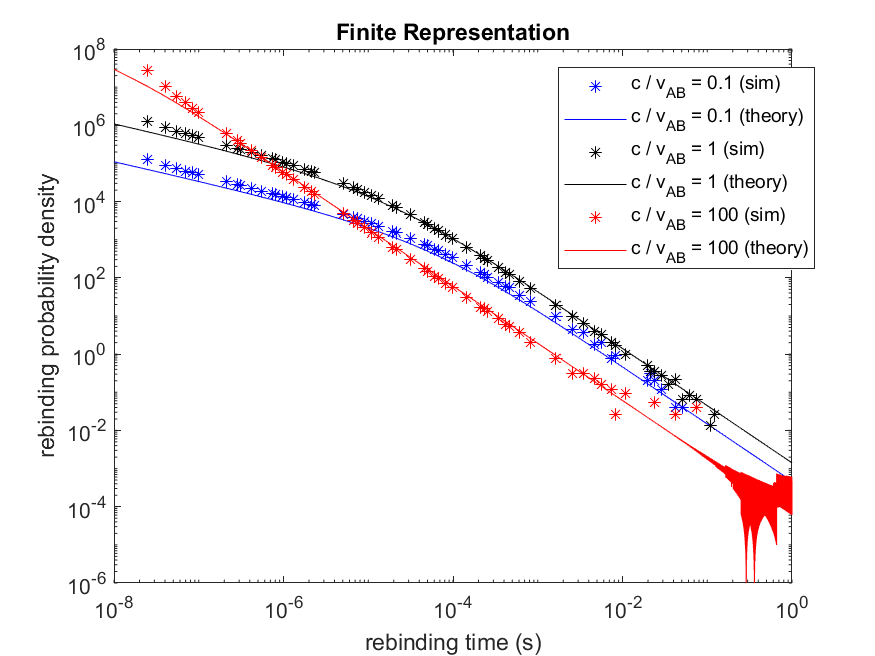}
  \caption{Rebinding time probability density from Noyes theory. We compare the theoretical curves in the finite particle representation with values computed from simulations at $c/\nu_{AB} = 0.1,\hspace{1mm} 1,\hspace{1mm}$ and $100$. Deviations from the theory at larger rebinding times are explained by the fact that more samples are required to characterize the probability densities at these limits than were drawn in our analysis. Simulation parameters were: $D_{A} = 1 \mu m^{2} s^{-1}$, $D_{B} = 1 \mu m^{2} s^{-1}$, $r_{0} = 0.01001\mu m$. At the largest times, numerical instabilities can appear in the Green's function computations, leading to the noise seen in the $c/\nu_{AB} = 100$ theoretical (theory) data. }
  \label{fig:rebindDensity}
\end{figure*}

\begin{figure*}[]
\includegraphics[width=1.00\textwidth]{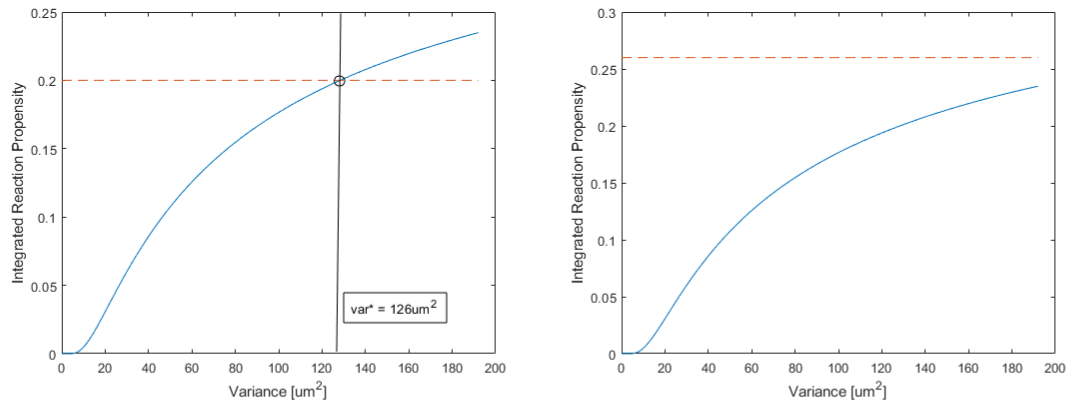}
\caption{Examples of successful and unsuccessful sampling of a biomolecular reaction in the point particle representation.  In both subfigures, the solid curve is the integrated reaction propensity associated with two reactants described by Gaussians with means separated by $8\mu m$. The left subfigure shows the successful sampling of a bimolecular reaction waiting time as there is a variance value (and thus, a waiting time) at which the integrated reaction propensity equals the exponentially distributed random number, $0.2$. In the right subfigure, the exponentially distributed random number is $0.26$, and so there is not sufficient integrated propensity for a reaction to occur. In the finite particle representation, the x-axis will represent the waiting time directly instead of variance.}
\label{fig:Combo__successful_and_unsuccessful}
\end{figure*}

Noyes theory is formulated in terms of the probability two molecules will re-collide (and potentially rebind) following a nonreactive encounter, therefore comparing the theoretical and simulated rebinding time probability densities is a useful test of our Gillespie-inspired wait time sampling procedure. Following Chew et al.~\cite{chew2018reaction}, we consider both the activation-limited and diffusion-influenced cases (Figure 2). These are distinguished by $c/\nu_{AB} < 1$ and $c/\nu_{AB} \geq 1$, respectively. See
%Section II.D from
Gillespie et al. \cite{gillespie2014small} for a derivation of this relation. The parameter $c$ is the boundary value parameter appearing in the finite particle propensity function and is related to the collision frequency $\nu_{AB} = 4\pi R(D_{A}+D_{B})$ between A and B molecules in a hypothetical nonreactive system as:
$$ c = \frac{P(reaction \hspace{1mm} |\hspace{1mm} encounter) \nu_{AB}}{g(r=R) \rho_{B}}, $$
where $g(r)$ is the particle pair correlation function in a liquid phase and $\rho_{B}$ is the relative density of B molecules. See
%Section III. and Eq.21 of
\cite{van2001diffusion}. Figure \ref{fig:rebindDensity} shows, for three values of $c/\nu_{AB}$, the theoretical density and the results of our simulations. We computed the integrated reaction propensity at 70,000,000 time points linearly spaced in the range [$10^{-8}$,1]. On the order of $10^9$ wait time samples were drawn for each of the three ratios and then aggregated into bins of width $w_{bin} = 5 \times 10^{-9}$s. Simulated probability densities for a subset of bins were computed as
$$ pdf(bin) = \frac{N_{bin}}{N_{total} \hspace{1mm} w_{bin}},$$
where $N_{bin}$ is the number of samples in the bin and $N_{total}$ is the total number of samples. As the total number of samples grows, the simulated values approach the theoretical density.

\begin{algorithm}[H] % MT: [H] is necessary for compatibility with revtex 4.2 document class
    \caption{Sampling Bimolecular Wait Times - Point Particle Representation}
    \label{alg:sbwt}
    \begin{algorithmic}[1] % MT: this [1] includes numbering
            \State (Pre-simulation) Define vector of variance values, $\mathbf{v} = [0,V_{max}]$
            \State (Pre-simulation) Define the curve $IntF(\mathbf{v} \hspace{1mm}|\hspace{1mm} d,R_{enc}^{2},c)$ as the cumulative sum of reaction propensity values along the points $\mathbf{v}$. \{$IntF(\mathbf{v} \hspace{1mm}|\hspace{1mm} d,R_{enc}^{2},c)$\} is then the set of integrated propensity curves at increasing d, computed once, before the simulation begins. If desired, further sets of curves can be precomputed for alternate values of $R_{enc}$ and $c$.
            \State (At run time) For reactant pair k = (A,B), select the appropriate curve, $IntF(\mathbf{v} \hspace{1mm}|\hspace{1mm} d_{k},R_{enc}^{2},c)$
            \State Evaluate $IntF(v_{t_{offset}})$, the integrated propensity to be discounted, at the variance value corresponding to $t_{offset}$, i.e., $6D_{b}t_{offset}$. 
            \State Set $v^{*} \leftarrow  argmin_{v} \hspace{1mm} IntF(\mathbf{v}) \geq ln(1/r_{k}) + IntF(v_{t_{offset}})$
            \State If $v^{*}$ exists, $t_{wait}$ is the solution to $v^{*} = 6D_{a}t_{wait} + 6D_{b}(t_{wait} + t_{offset})$
            \State Else, no reaction is sampled. Update particle positions.
    \end{algorithmic}
\end{algorithm}

\section*{Sampling Bimolecular Reaction Locations}

\label{Sampling Bimolecular Reaction Locations}

Again we make use of the labels A and B for the specific molecules undergoing the next association reaction. At this time, the spatial region available for the reaction consists of the intersection of the diffusion spheres bounding their independent Gaussian probability distributions. In order to correctly sample from this region, henceforth called the \textit{overlap volume} (OV), we first introduce the concept of equiprobable rings.

\subsection*{Equiprobability Rings}

The line AB connecting the initial known positions of A and B defines an axis of symmetry in the sense that within the OV there exist rings centered on this axis, whose points are equidistant from A and equidistant from B. The rings are therefore sets of equiprobability points from which molecule positions might be sampled. Each ring is uniquely defined by two numbers: the magnitude, $r_{A}$, of any vector from the initial position of A to a point on the ring, and the CCW angle, $\theta_{A}$, between the vector and the line AB. After sampling $(r_{A},\theta_{A})$, we choose the reaction location uniformly at random from on the ring.\\
The joint probability density describing $(r_{A},\theta_{A})$ can be factored as $p(r_{A}|t)$ and the conditional probability $p(\theta_{A}|r_{A},t)$, which suggests a sequential sampling procedure. First determine $r_{A}$ and then use it to determine $\theta_{A}$. 

\subsection*{Diffusion Sphere Overlap Volume}

\begin{figure}[h]
  \includegraphics[width=0.8\linewidth]{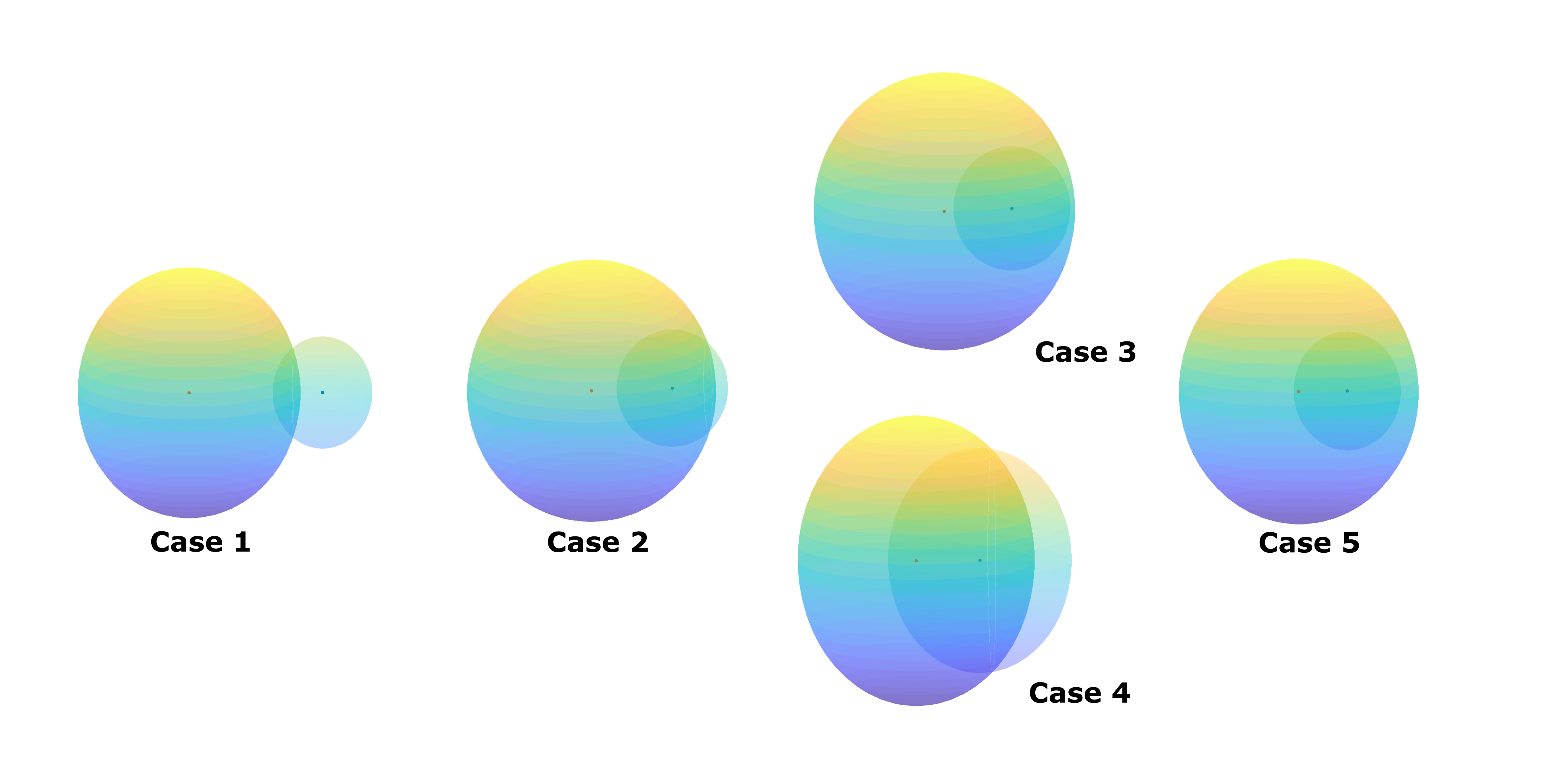}
  \centering
  \caption{Cases of potential overlap of diffusion spheres in the process of sampling waiting time to a biomolecular reaction.  Shown are the diffusion sphere intersections at increasing time points. It is assumed here that $D_{B} > D_{A}$. Case 1: The OV contains neither $\mu_{A}$ nor $\mu_{B}$. Case 2: The OV contains $\mu_{A}$ only, and is not identical to either diffusion sphere. Case 3: The OV contains $\mu_{A}$ only, and is identical to the diffusion sphere of A. Case 4: The OV contains $\mu_{A}$ and $\mu_{B}$, but is not identical to either diffusion sphere. Case 5: The OV contains $\mu_{A}$ and $\mu_{B}$, and is identical to the diffusion sphere of A.}
  \label{fig:diffusion_intersection_trajectories}
\end{figure}

While the OV grows continuously due to diffusion, for the purpose of location sampling at a given time we have found it useful to classify it into one of five distinct cases. These cases are not inherently meaningfully different in the theory behind them, but provide a convenient way of describing the evolution of the OV as well as distinguishing the integration regions involved in sampling, $r_{A}$ and $\theta_{A}$, for purposes of clearer exposition. For example, if the OV is identical to the diffusion sphere of A (as in cases 3 and 5), $\theta_{A}$ may take on any value in [$0,2\pi$], however if the OV has an irregular shape, certain angles may be prohibited. Figure \ref{fig:diffusion_intersection_trajectories} visualizes the two trajectories possible for the OV. The first trajectory applies when $D_{B} > 4D_{A}$ and passes through cases 1, 2, 3 and 5. The second trajectory applies when $D_{A} < D_{B} < 4D_{A}$ and passes through cases 1, 2, 4 and 5. Given the current system time $t$, the waiting time until the next reaction of A and B, $t_{wait}$, and the system time at which the state B was last updated, we can infer $t_{A-elapsed}$ and $t_{B-elapsed}$, the durations during which each had been diffusing before the reaction, which includes the waiting time to the reaction. Using $t_{A-elapsed}$ and $t_{B-elapsed}$ to define the diffusion spheres at the moment the molecules react, we can infer the OV case.

\begin{widetext}
\begin{figure*}[!]
  \includegraphics[width=\textwidth]{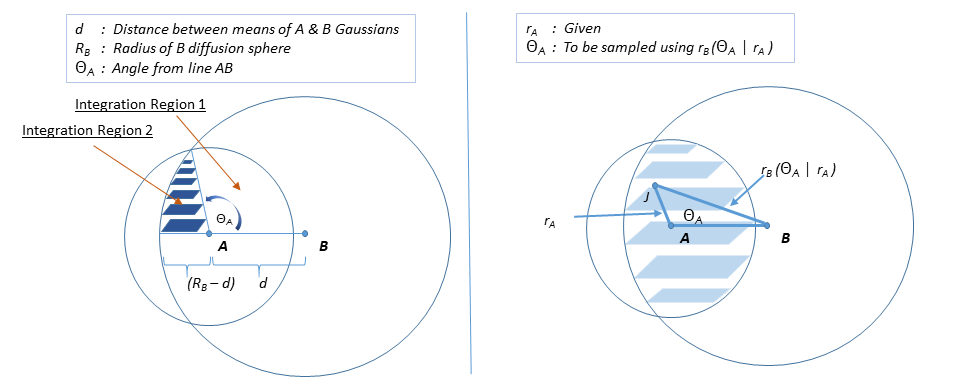}
  \caption{(\textbf{Left}) Visualizing the regions of integration for $w(r_{A})$ in Case 2. (\textbf{Right}) Visualizing $\theta_{A}$, $r_{A}$, and $r_{B}(\theta_{A})$ in Case 2. The equiprobability ring passes through point $J$, perpendicular to the plane of the page.}
  \label{fig:Case2_Sampling_rA_thetaA_Diagram_Combo}
\end{figure*}
\end{widetext}

Case 2 begins when the radius of the faster diffusing particle (here, B) is equal to $d$, the distance between the Gaussian means of A and B. This radius can be computed as $R_{B}(t) = n_{sigma}\sqrt{6D_{B}t}$, where $n_{sigma}$ is the number of standard deviations bounded by the sphere. See Fig.~\ref{fig:Case2_Sampling_rA_thetaA_Diagram_Combo} for an illustration of the integration variables in Case 2. The starting time is given by

\begin{equation}
t_{start-2} = \frac{d^2}{6 D_{B} n_{sigma}^{2}}
\end{equation}

\noindent
Starting times for cases 3-5 are calculated as follows:

\begin{center}
\begin{tabular}{ |c|c|c|c| } 
\hline
Path 1 & Case 3 Start & Case 5 Start \\
\hline
\multirow{3}{7em}{$D_{B} > 4D_{A}$} & $d > R_{A}(t)$ & $d = R_{A}(t)$ \\ 
& $R_{A}(t) + d = R_{B}(t)$ & $R_{A}(t) + d < R_{B}(t)$ \\
& $t_{start-3} = t_{\gamma} $ & $t_{start-5} = t_{\sim \gamma} $\\
\hline
%\end{tabular}
%\end{center}

%\begin{center}
%\begin{tabular}{ |c|c|c|c| } 
\hline
Path 2 & Case 4 Start & Case 5 Start \\
\hline
\multirow{2}{8em}{$D_{A} < D_{B} < 4D_{A}$} & $d = R_{A}(t)$ & $d < R_{A}(t)$ \\ 
& $R_{A}(t) + d > R_{B}(t)$ & $R_{A}(t) + d = R_{B}(t)$ \\
& $t_{start-4} = t_{\sim \gamma} $ & $t_{start-5} = t_{\gamma} $\\
\hline
\end{tabular}
\end{center}

%\noindent
where 

\begin{multline}
    t_{\gamma} = \frac{1}{{6 n_{sigma}^{2} (D_{A} - D_{B})^{2}}} (2 D_{A}^{2} n_{sigma}^{2} \gamma + 2 D_{B}^{2} n_{sigma}^{2}  \gamma \\- 4 D_{A} D_{B} n_{sigma}^{2}  \gamma + D_{A} d^{2} + D_{B} d^{2}),
\end{multline}

$t_{\sim \gamma} = \frac{d^2}{6 D_{A} n_{sigma}^{2}}$,
and $\gamma = \sqrt{\frac{D_{A} D_{B} d^{4}}{n_{sigma}^{4} (D_{A} - D_{B})^{4}}}$.

% % ----------------------------------------
% % ---------------- CASE 1 ----------------
% % ----------------------------------------
\subsection*{\textbf{Case 1}}

\subsubsection*{Sampling $r_{A}$}

In order to sample $r_{A}$ correctly, we re-weight the probability density in the OV, i.e., compute a posterior probability.  Define $h_{ring}(\theta_{A})$ as the radius of the ring whose points are at distance $r_{A}$ and for which the top most point defines a line with A at angle $\theta_{A}$. The circumference of this ring is $2\pi h_{ring}(\theta_{A})$. Integrating this circumference over the available $\theta_{A}$ range allows us to determine the size of the set of points at distance $r_{A}$. 
\begin{equation}
p_{reweighted}(r_{A},t) = w(r_{A})*p(r_{A},t)
\end{equation}
with
$$ w(r_{A}) = \frac{[Total Probability-at-r_{A}]}{\int_{OV} dr \Big( \hspace{1mm} [Total Probability-at-r]*p(r,t) \Big)},$$

\begin{equation}
\int p_{reweighted}(r_{A},t) dr_{A} = \int w(r_{A})p(r_{A},t) = 1,
\end{equation}
and
\begin{equation}
p(r,t) = \frac{1}{\sqrt{12\pi D_{A}t}} exp(-r^2/12D_{A}t)
\end{equation}

\begin{multline}
w(r_{A}) = \frac{\int_{0}^{\theta_{max}(r_{A})} d\theta_{A} 2\pi h_{ring}(\theta_{A})} {\int_{r_{lb}}^{r_{ub}} dr \Big[ \big( \int_{0}^{\theta_{max}(r)} d\theta(r) 2\pi r \sin(\theta) \big) p(r,t) \Big]}\\
= \frac{r_{A}\big( cos(\theta_{max}(r_{A})) - cos(0) \big)}{\int_{r_{lb}}^{r_{ub}} dr \Big[ r \big(cos(\theta_{max}(r)) - cos(0)\big) p(r,t) \Big]}\\
= \frac{r_{A} \left(\frac{r_{A}^{2} + d^{2} - R_{B}^{2}}{2r_{A}d} -1 \right)}{ \Big[ term1 + term2   \Big]}
\end{multline} 

\begin{multline*}
term1 = \frac{1}{4d} (d^2+6D_{A}t-R_{B}^{2})\big[ erf\big( r_{ub}/\sqrt{12D_{A}t}\big) \\- erf\big( r_{lb}/\sqrt{12D_{A}t}\big) \big]
\end{multline*}

\begin{multline*}
term2 = \frac{1}{\sqrt{12\pi D_{A}t}}6D_{A}t \big[ (r_{lb}-2d)exp(-r_{lb}^{2}/12D_{A}t) \\- (r_{ub}-2d)exp(-r_{ub}^{2}/12D_{A}t) \big]
\end{multline*}

The upper limit of integration, $\theta_{A-max}$, is calculated by considering the triangle defined by the three points: $A$, $B$, $I$. The base (AB) length is $d$. The side $BI$ has length $R_{B}$ since $I$ is the point at which ($r_{A},\theta_{A}$) intersects the OV, i.e., a point on the B diffusion sphere. The remaining side length is $r_{A}$. From the law of cosines, $\theta_{A-max}$ is calculated in terms of the side lengths.
\begin{equation}
\theta_{A-max}(r) = cos^{-1}\left(\frac{r^{2} + d^{2} - R_{B}^{2}}{2rd}\right)
\end{equation}

The lower and upper bounds, $r_{lb}$ and $r_{ub}$, on $r_{A}$ defining the OV are 
$[(d - R_{B}),R_{A}]$.

\subsubsection*{Sampling $\theta_{A}|r_{A},t$}

The tuple $(\theta_{A},r_{A})$ uniquely defines a ring of equiprobability points within the OV from which a single reaction location can be chosen uniformly at random. Thus, the probability with which a given $\theta_{A}$ is sampled should be proportional to the size of the corresponding ring. 

Consider the triangle defined by the points $A$, $B$, $J$ where $J$ is a point in the OV at $(\theta_{A},r_{A})$. The length of side $BJ$ is $r_{B}(\theta_{A})$ and can be computed with the Law of Cosines. The height of this triangle, $h_{ring}$, is again the radius of the ring passing through point $J$. 

\begin{equation*}
p(\theta_{A}|r_{A},t) = p(r_{B}(\theta_{A})|t)*RingCircumference
\end{equation*}
\begin{equation}
p(\theta_{A}|r_{A},t) = \frac{1}{\sqrt{12\pi D_{B}t}} \hspace{1mm} exp\left(-\frac{r_{B}(\theta_{A})^{2}}{12 D_{B}t}\right) * 2\pi h_{ring}
\end{equation}

\begin{equation}
r_{B}^{2}(\theta_{A}) = r_{A}^{2} + d^{2} - 2r_{A}d \hspace{1mm} cos(\theta_{A})
\end{equation}
\begin{equation}
h_{ring} = r_{A} \hspace{1mm} sin(\theta_{A})
\end{equation}
$$\theta_{A} \in [0,\theta_{A-max}]$$
\iffalse
\begin{multline*}
h_{ring} = \frac{2*Area}{base} = \frac{\sqrt{s(s-r_{A})(s-r_{B})(s-d)}}{d}\\ = \frac{\sqrt{(\frac{r_{A}+r_{B}+d}{2})((\frac{r_{A}+r_{B}+d}{2})-r_{A})((\frac{r_{A}+r_{B}+d}{2})-r_{B})((\frac{r_{A}+r_{B}+d}{2})-d)}}{d}
\end{multline*}
\fi

% % ----------------------------------------
% % ---------------- CASE 2 ----------------
% % ----------------------------------------
\subsection*{\textbf{Case 2}}

\subsubsection*{Sampling $r_{A}$}
\begin{multline}
w(r_{A}) = \int_{0}^{\theta_{A-max}(r_{A})} d\theta_{A} 2\pi r_{A}sin(\theta_{A}) \hspace{2mm}*\\
\Big(\int_{0}^{R_{B}-d} dr \big[ \int_{0}^{\theta_{max}(r)} d\theta(r) 2\pi r \sin(\theta) \big]*p(r,t)\\
+ \int_{R_{B}-d}^{R_{A}} dr \big[ \int_{0}^{\theta_{max}(r)} d\theta(r) 2\pi r \sin(\theta) \big]*p(r,t)\Big)^{-1}
\end{multline}

\begin{equation}
\theta_{A-max}(r) = cos^{-1}\Big(\frac{max[r^{2},(R_{B}-d)^{2}] + d^2 - R_{B}^{2}}{2d \hspace{1mm} max[r,(R_{B}-d)]} \Big)
\end{equation}

Figure \ref{fig:Case2_Sampling_rA_thetaA_Diagram_Combo} provides a visual description of the relevant Case 2 variables. Variables for the other cases are defined similarly. For any $r_{A}$ less than or equal to $(R_{B}-d)$, the full angular range of region 2 is available, i.e., $\theta \in (0,\pi)$. As $r_{A}$ increases from $(R_{B}-d)$ to $R_{A}$, the available positions within region 2 decrease to 0. We capture this dependence with the angle integration limits, ($0,\theta_{A-max}$), where $\theta_{A-max} = \pi$ for $r_{A} \leq (R_{B}-d)$. The logic behind the form of $w(r_{A})$ is analogous to case 1, however.

\subsubsection*{Sampling $\theta_{A}|r_{A},t$}

Sampling here is analogous to case 1, with updates to the available angle ranges for a given $r_{A}$.
\begin{equation}
p(\theta_{A}|r_{A},t) = \frac{1}{\sqrt{12\pi D_{B}t}} \hspace{1mm} exp\left(-\frac{r_{B}(\theta_{A})^{2}}{12 D_{B}t}\right) * 2\pi h_{ring}
\end{equation}
With $r_{B}^{2}(\theta_{A}) = r_{A}^{2} + d^{2} - 2r_{A}d \hspace{1mm} cos(\theta_{A})$, $h_{ring} = r_{A} \hspace{1mm} sin(\theta_{A})$, and $\theta_{A} \in [0,\theta_{A-max}]$. 

% % ----------------------------------------
% % ---------------- CASE 2 ----------------
% % ----------------------------------------

% % ----------------------------------------
% % ---------------- CASE 3 ----------------
% % ----------------------------------------
%\begin{widetext}
\subsection*{\textbf{Case 3}}
\subsubsection*{Sampling $r_{A}$}

In this case, the full range in $r_{A}$ ($\in [0,R_{A}]$) is available. Therefore, no re-weighting of probabilities is needed.
\begin{equation}
p(r_{A}|t) = \frac{1}{\sqrt{12\pi D_{A}t}} \hspace{1mm} exp\left(-\frac{r_{A}^{2}}{12 D_{A}t}\right)
\end{equation}

\subsubsection*{Sampling $\theta_{A}|r_{A},t$}
Sampling here is analogous to case 1, but with the full range of angles available.
\begin{equation}
p(\theta_{A}|r_{A},t) = \frac{1}{\sqrt{12\pi D_{B}t}} \hspace{1mm} exp\left(-\frac{r_{B}(\theta_{A})^{2}}{12 D_{B}t}\right) * 2\pi h_{ring}
\end{equation}
With $r_{B}^{2}(\theta_{A}) = r_{A}^{2} + d^{2} - 2r_{A}d \hspace{1mm} cos(\theta_{A})$, $h_{ring} = r_{A} \hspace{1mm} sin(\theta_{A})$, and $\theta_{A} \in [0,\pi]$.
%\end{widetext}

% % ----------------------------------------
% % ---------------- CASE 3 ----------------
% % ----------------------------------------
% % ----------------------------------------
% % ---------------- CASE 4 ----------------
% % ----------------------------------------
\subsection*{\textbf{Case 4}}
\subsubsection*{Sampling $r_{A}$}
Sampling here is analogous to case 2.

\begin{multline}
w(r_{A}) = \int_{0}^{\theta_{A-max}(r_{A})} d\theta_{A} 2\pi r_{A}sin(\theta_{A}) \hspace{2mm}*\\
\Big(\int_{0}^{R_{B}-d} dr \big[ \int_{0}^{\theta_{max}(r)} d\theta(r) 2\pi r \sin(\theta) \big]*p(r,t)\\
+ \int_{R_{B}-d}^{R_{A}} dr \big[ \int_{0}^{\theta_{max}(r)} d\theta(r) 2\pi r \sin(\theta) \big]*p(r,t)\Big)^{-1}
\end{multline}

\begin{equation}
\theta_{A-max}(r) = cos^{-1}\Big(\frac{max[r^{2},(R_{B}-d)^{2}] + d^2 - R_{B}^{2}}{2d \hspace{1mm} max[r,(R_{B}-d)]} \Big)
\end{equation}

\subsubsection*{Sampling $\theta_{A}|r_{A},t$}
Sampling here is also analgous to case 2.
\begin{equation}
p(\theta_{A}|r_{A},t) = \frac{1}{\sqrt{12 \pi D_{B}t}} \hspace{1mm} exp\left(-\frac{r_{B}(\theta_{A})^{2}}{12 D_{B}t}\right) * 2\pi h_{ring}
\end{equation}
With $r_{B}^{2}(\theta_{A}) = r_{A}^{2} + d^{2} - 2r_{A}d \hspace{1mm} cos(\theta_{A})$, $h_{ring} = r_{A} \hspace{1mm} sin(\theta_{A})$, and $\theta_{A} \in [0,\theta_{A-max}]$. 

% % ----------------------------------------
% % ---------------- CASE 5 ----------------
% % ----------------------------------------
\subsection*{\textbf{Case 5}}
\subsubsection*{Sampling $r_{A}$}
In this case, the full range in $r_{A}$ ($\in [0,R_{A}]$) is available. Therefore, no re-weighting of probabilities is needed.
\begin{equation}
p(r_{A}|t) = \frac{1}{\sqrt{12\pi D_{A}t}} \hspace{1mm} exp(-\frac{r_{A}^{2}}{12 D_{A}t})
\end{equation}

\subsubsection*{Sampling $\theta_{A}|r_{A},t$}
Sampling here is analogous to case 1, but with the full range of angles available.
\begin{equation}
p(\theta_{A}|r_{A},t) = \frac{1}{\sqrt{12 \pi D_{B}t}} \hspace{1mm} exp\left(-\frac{r_{B}(\theta_{A})^{2}}{12 D_{B}t}\right) * 2\pi h_{ring}
\end{equation}
With $r_{B}^{2}(\theta_{A}) = r_{A}^{2} + d^{2} - 2r_{A}d \hspace{1mm} cos(\theta_{A})$, $h_{ring} = r_{A} \hspace{1mm} sin(\theta_{A})$, and $\theta_{A} \in [0,\pi]$.

\subsection*{Determining Bimolecular Reaction Locations by Rejection Sampling}

Because PDFs in each case may be complicated functions, we cannot always sample from them directly. Instead, we first draw a sample of our variable x (i.e. $r_{A}$ or $\theta_{A}$) uniformly from its feasible range. In order to determine whether this sample is accepted or rejected, we utilize an envelope function, $Q(x)$ whose probability density at all feasible points is at least as great as that of the PDF from which we want an observation.  The sample $x$ is accepted if $q(x)$ drawn uniformly from $[0,Q(x)]$ is less than $p(x)$. 

One potential issue is that volume exclusion should prevent sampled locations from leading to particle overlap. We define two molecules to be overlapping if the distance between their Gaussian means is less than $R$, the minimum allowed separation, \textit{and} neither molecule has been diffusing for longer than $R^{2}/6D$. In the event overlap is detected, a new location is sampled. This procedure for handling volume exclusion in location sampling is not highly optimized in the demonstration implementation presented here, and can significantly impact the run time as the particle density increases.

%We use a uniform distribution as the envelope function with the maximum defined to be larger than the true PDF for all feasible inputs. Future work will focus on optimizing the construction of the envelope function in order to minimize the number of rejected samples.

\subsection*{Simulation Boundaries}
\begin{figure*}[!]
  \includegraphics[width=\textwidth]{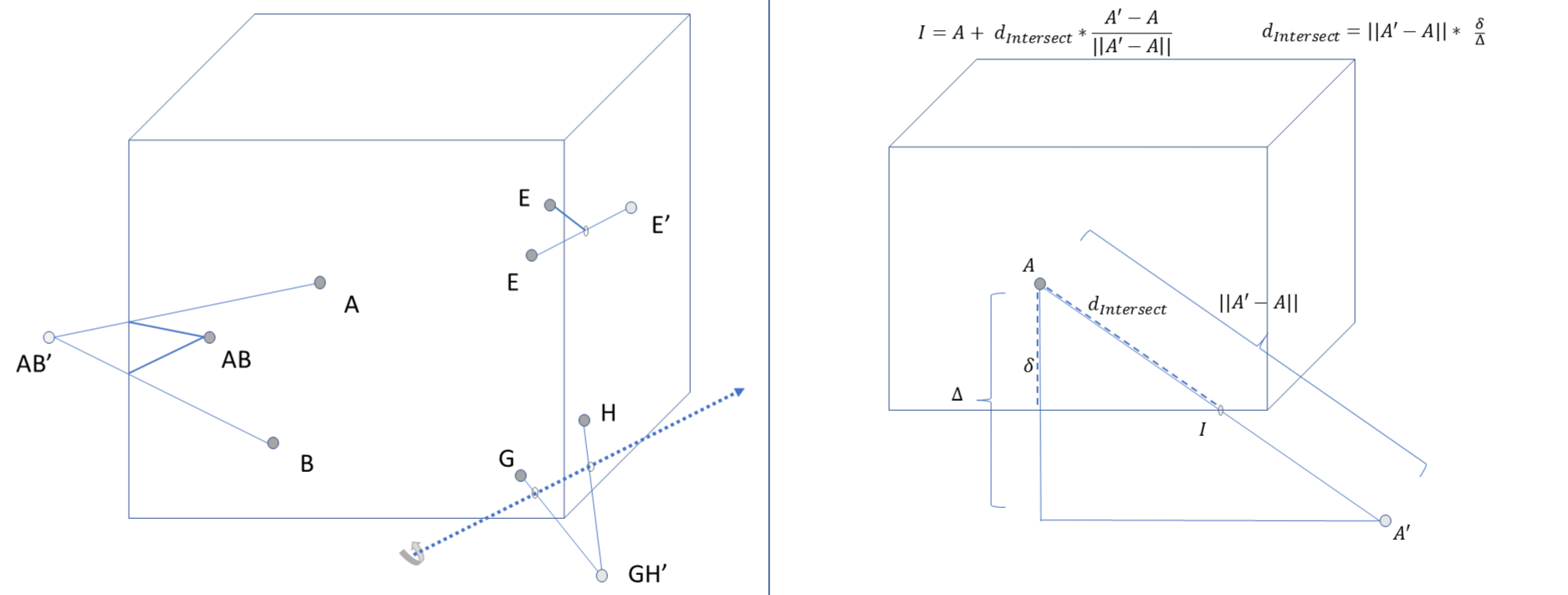}
  \caption{(\textbf{Left}) Shown are applications of the reflective boundary condition after a position-only-update event (e.g. E) or after a bimolecular reaction event (e.g. A\&B, G\&H). (\textit{Single Reflection}) In the bimolecular case, we reflect about an axis defined by the two intersection points of the lines connecting the reactants with the product, and the boundary. This ensures that the distances traveled by both particles remains the same. When these lines exit the simulation box through the same face (e.g. A\&B), the reflection axis is parallel to the face. When the lines exit though different faces (e.g. G\&H), the axis must be computed and the reflection can be implemented with the Rodrigues rotation formula in the appropriate reference frame. (\textit{Multiple Reflection}) Depending on the location of the reactants and the distances they travel, the post-reflection location may end up outside a different boundary, though to a lesser extent. We simply need to update the reactant positions to be the boundary intersection point(s) and reapply the reflection procedure. In principle, this procedure works for any simulation volume, including those with curved boundaries. (\textbf{Right}) For a cubic simulation volume, we determine through which face (and at what point) a reactant (A) first passed if it is found outside the simulation volume. In this case, the pre-reflection location A' exceeds the simulation volume along more than 1 dimension which means it is necessary to compute $d_{Intersect}$ for each 2d plane exceeded by A', and then compute the intersection point, I, for the face with minimum $d_{Intersect}$.}
  \label{fig:implementing_reflections}
\end{figure*}

Figure \ref{fig:implementing_reflections} illustrates our method for ensuring all particles remain within the simulation volume. We treat this volume as a cube bounded by planes about which a particle may be reflected if its initially sampled position exceeds the plane. The algorithm samples an unconstrained reaction location and the displacements for both particles are noted. Next, assume the reaction location happens to be outside the simulation volume. Each molecule can be considered to have travelled along a linear path from its initial location to the reaction location, with one piece of the path within the simulation volume and one piece outside. Because the unconstrained spatial probability densities describe radial displacements from either particle's initially known location, application of reflective boundary conditions need only guarantee both particles' piecewise linear paths each sum to the noted displacements, and terminate within the simulation volume.

This procedure is strictly correct only if the wait time sampling, i.e., computing the integrated reaction propensities, is correct. The point (finite) particle reaction propensities described in this paper do not take into account the boundaries of the simulation volume. Error is therefore introduced in wait time sampling for molecules diffusing long enough to encounter a boundary. However, given a wait time $t$, reaction location sampling depends only on the possible net displacements of either particle after diffusing for $t$ units of time. We can therefore assume free diffusion to sample the location and then use our reflecting procedure if necessary.

\iffalse
One potential issue is that within the bounded simulation volume, the reaction propensity never decays to zero. Instead, it eventually reaches a plateau equal to the well-mixed approximation. Therefore, we should limit the diffusion time such that the unconstrained particles might diffuse to a distance greater than a boundary, but not so long that the well-mixed assumption applies.
\fi

\iffalse
\begin{figure}[h]
%  \includegraphics[width=\textwidth]{which_boundary_exit_first.png}
  \caption{Illustration of method to determine through which face (and at what point) a reactant (A) first passes if, after event execution, it is found outside the simulation volume. In this case, the pre-reflection location A' exceeds the simulation volume along more than 1 dimension which means it is necessary to compute $d_{Intersect}$ for each 2d plane exceeded by A', and then compute the intersection point, I, for the face with minimum $d_{Intersect}$.  }
  \label{fig:which_boundary_exit_first}
\end{figure}
\fi

\begin{figure*}[!]
  \includegraphics[width=\textwidth]{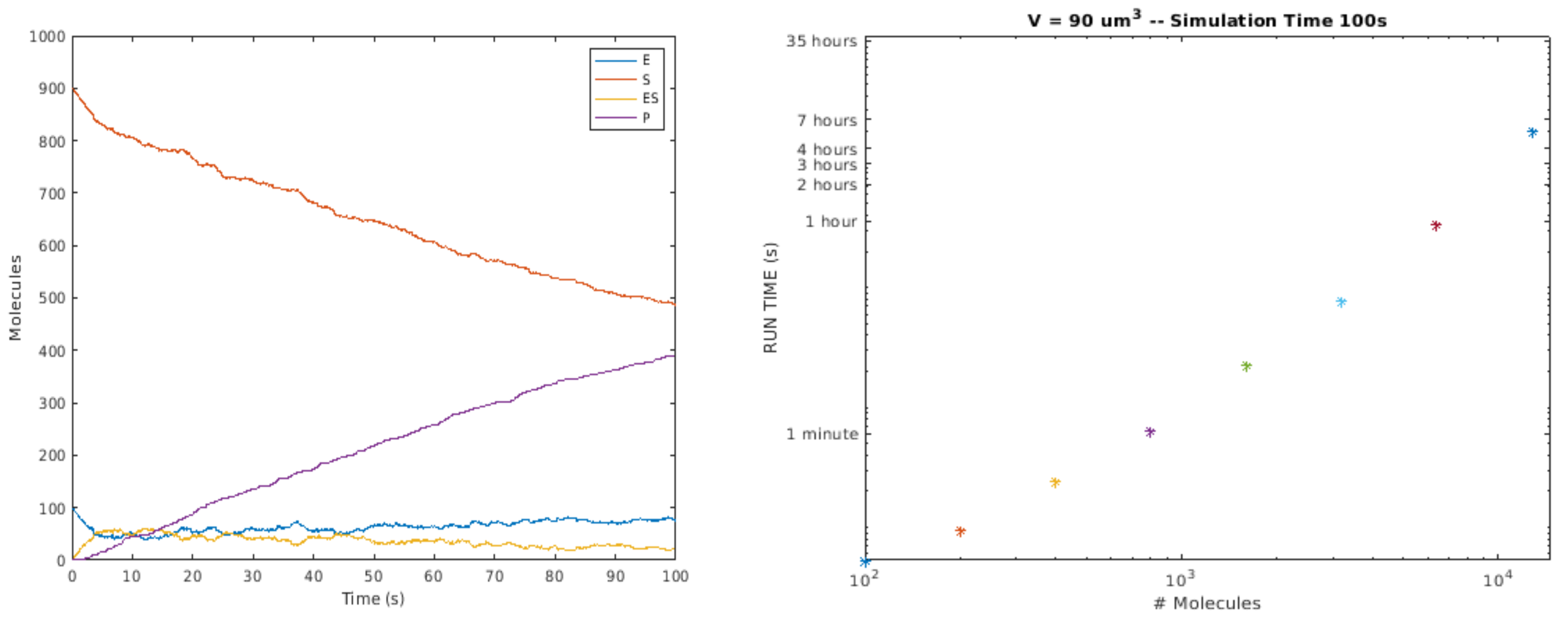}
  \caption{ Point particle representation. (\textbf{Left}) Time evolution of 1000 molecules in the Michaelis-Menten model with DESSA-CS. Unimolecular rate constant $k_{uni} = 0.1s^{-1}$ (governing $ES \Rightarrow E + S$ and $ES \Rightarrow E + P$) and diffusion coefficient $D = 1 \mu m^{2} s^{-1}$ are taken from Figure 5 of Chew et al. \cite{chew2018reaction}. In order to reproduce similar dynamics, we chose the point-particle reaction propensity constant
  %rate constant 
  $c = 2.5*10^{7}s^{-1}$ (with $R_{enc}^{2} = 0.01^{2}\mu m$). (\textbf{Right}) The run time for the model increases roughly linearly in log space with the number of molecules - [100,200,400,800,1600,3200,6400,12800] at a fixed volume of 90 $\mu m^3$.}
  \label{fig:MM_reaction_plot}
\end{figure*}

\begin{figure*}[!]
  \includegraphics[width=\textwidth]{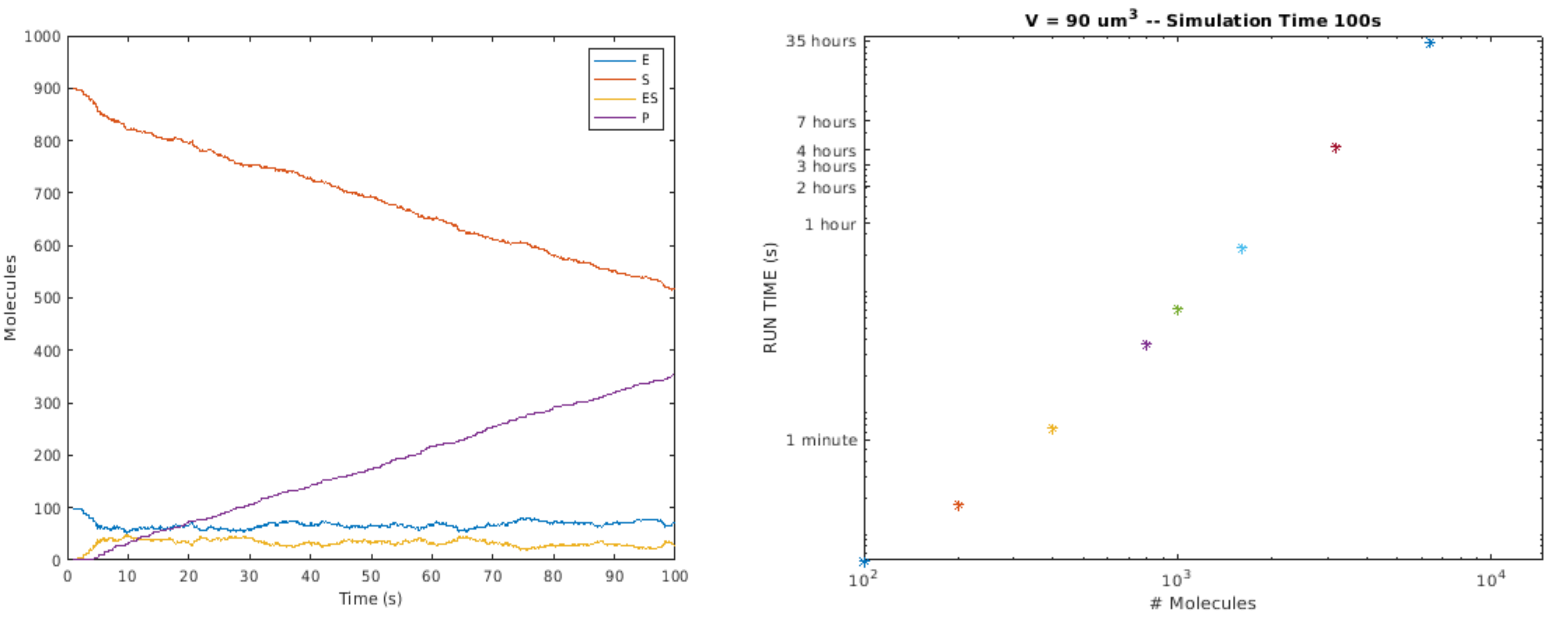}
  \caption{ Finite particle representation. (\textbf{Left}) Time evolution of 1000 molecules in the Michaelis-Menten model with DESSA-CS. Unimolecular rate constant $k_{uni} = 0.1s^{-1}$ (governing $ES \Rightarrow E + S$ and $ES \Rightarrow E + P$), diffusion coefficient $D = 1 \mu m^{2} s^{-1}$, boundary condition parameter $c = 1.2*10^{3}s^{-1}$, and $R_{enc} = 0.01\mu m$.  (\textbf{Right}) The run time for the model increases roughly linearly in log space with the number of molecules - [100,200,400,800,1600,3200,6400,12000] at a fixed volume of 90 $\mu m^3$. }
  \label{fig:MM_reaction_plot_finiteRep}
\end{figure*}

% ___________________________________________________________________
%               1st Methods Comparison Table Begins Here
% ___________________________________________________________________
\setlength{\arrayrulewidth}{1mm}
\setlength{\tabcolsep}{5pt}
\renewcommand{\arraystretch}{2.0}
\begin{table*} 
{%\rowcolors{5}{green!90!yellow!10}{green!70!yellow!30}
%\begin{tabular}{ |p{0.6cm}|p{0.5cm}|p{1cm}|p{1cm}|p{1cm}|p{1cm}|p{1cm}|  }
\begin{tabular}{ |c|c|c|c|c|c|  }
\hline
\multicolumn{6}{|c|}{1000 Molecules, 100s Simulation Time} \\
\hline
\multicolumn{6}{|c|}{Local Workstation: Ubuntu 14.04 LTS, 128 GB memory, Intel Xeon E5-2630 2.40GHz} \\
\hline
\multicolumn{6}{|c|}{\cite{chew2018reaction} Workstation: Ubuntu 16.04 LTS, 48 GB memory, Intel Xeon X5680 3.33GHz} \\
\hline
\underline{Software} & \underline{Run Time}& \underline{Sim Parameters}& \underline{Boundary Type}& \underline{Space / Time Steps}& \underline{Workstation}\\
\hline
DESSA-CS & 100s & $r=0nm$, $l_{B}=0.1$ &reflective &off-lattice / sampled &local \\
DESSA-CS & 728s & $r=10nm$, $l_{B}=0.1$ &reflective &off-lattice / sampled &local \\
eGFRD & 10,561s & $r=10nm$ &periodic &off-lattice / variable & local\\
eGFRD & 2,412s & $r=1nm$ &periodic &off-lattice / variable &\cite{chew2018reaction}\\
eGFRD & 3,246s & $r=10nm$ &periodic &off-lattice / variable &\cite{chew2018reaction}\\
Smoldyn & 20s & $\Delta t=1ms$ &periodic &off-lattice / fixed &\cite{chew2018reaction}\\
Smoldyn & 298s & $\Delta t=67\mu s$ &periodic &off-lattice / fixed &\cite{chew2018reaction}\\
Spaciocyte MLM & 13s & $\Delta t=1ms$, $r=38.73nm$ &periodic &spatial lattice / fixed &\cite{chew2018reaction}\\
Spaciocyte MLM & 276s & $\Delta t=67\mu s$, $r=10nm$ &periodic &spatial lattice / fixed &\cite{chew2018reaction}\\
\hline
\end{tabular}
}
\caption{Method Comparison on Updated Benchmark from Chew et al. \cite{chew2018reaction}. Diffusion coefficients are $1 \mu m^{2}s^{-1}$. The local eGFRD simulation was run using the open source simulation environment E-Cell version 4 \cite{ecell4}. DESSA-CS parameter $l_{B}$ describes, as a percentage of the container length, how far a molecule's diffusion sphere may extend beyond the container during location sampling.}
\label{table:updated_benchmark}
\end{table*}

\iffalse
% ___________________________________________________________________
%               2nd Methods Comparison Table Begins Here
% ___________________________________________________________________
\begin{table*} 
{%\rowcolors{4}{green!90!yellow!10}{green!70!yellow!30}
%\begin{tabular}{ |p{0.6cm}|p{0.5cm}|p{1cm}|p{1cm}|p{1cm}|p{1cm}|p{1cm}|  }
\begin{tabular}{ |c|c|c|c|c|c|  }
\hline
\multicolumn{6}{|c|}{10,000 Molecules, 10s Simulation Time} \\
\hline
\multicolumn{6}{|c|}{Local Workstation: Ubuntu 14.04 LTS, 128 GB memory, Intel Xeon E5-2630 2.40GHz} \\
\hline
\underline{Software} & \underline{Run Time}& \underline{Sim Parameters}& \underline{Boundary Type}& \underline{Space / Time Steps}& \underline{Workstation}\\
\hline
DESSA-CS & 548s & max diffusion 10s &reflective &off-lattice / sampled &local \\
Smoldyn & 341s & $\Delta t=67\mu s$ &reflective &off-lattice / fixed &local\\
\hline
\end{tabular}
}
\caption{Method Comparison on Original Benchmark from Andrews. For the Smoldyn test, we ran the file \path{examples/S8_reactions/benchmark/benchmark.txt} updated to use reflective boundaries, from the Github repository \cite{andrews_GITHUB_Smoldyn}. Diffusion coefficients are $10 \mu m^{2}s^{-1}$. }
\label{table:updated_benchmark2}
\end{table*}
\fi

\section{Results}

\subsection*{Application: Michaelis-Menten}
\begin{figure}[h!]
  \includegraphics[width=.48\textwidth]{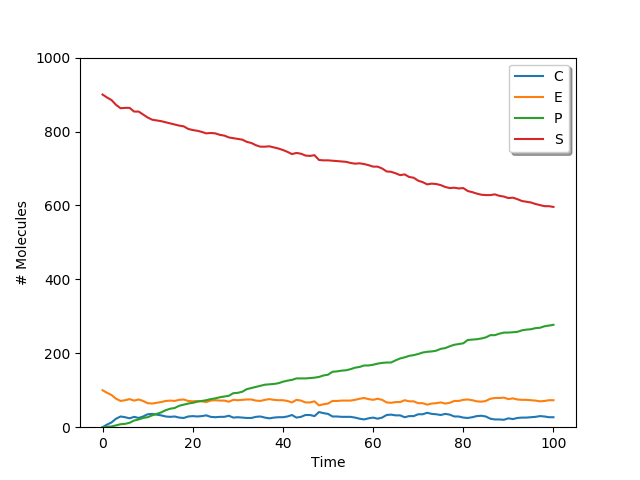}
  \caption{Time evolution of 1000 molecules in the Michaelis-Menten model with eGFRD in the E-Cell v4 environment. Unimolecular rate constant $k_{uni} = 0.1s^{-1}$ (governing $ES \Rightarrow E + S$ and $ES \Rightarrow E + P$), diffusion coefficient $D = 1 \mu m^{2} s^{-1}$, intrinsic bimolecular rate constant $k_{bimol} = 1*10^{-2}s^{-1}$, and particle radius $r = 0.01\mu m$ are taken from Figure 5 of Chew et al. \cite{chew2018reaction}}
  \label{fig:MM_100s__eGFRD__e-cell4}
\end{figure}

We applied DESSA-CS to the well known Michaelis-Menten enzymatic reaction system within a $90 \mu m^{3}$ volume. The original benchmark was developed by Andrews \cite{andrews2010detailed} and updated by Chew et al. \cite{chew2018reaction} to account for the extreme run time demands of eGFRD. 
Figures \ref{fig:MM_reaction_plot} and \ref{fig:MM_reaction_plot_finiteRep} display our results for the updated benchmark, displaying the population dynamics for molecular species E, S, ES and P, which obey the binding rules $E + S \Leftrightarrow ES \Rightarrow P.$  Figure \ref{fig:MM_100s__eGFRD__e-cell4} shows comparable results for eGFRD.  Note that the different models approximate and parameterize the physical system in different ways and so it is not possible to conduct perfectly equivalent simulations by each method, but we have chosen parameter values so as to approximate the same physical conditions in each method as closely as they allow in our choice of rate constants.
Simulation run times for the point particle and finite particle representations respectively were 20 seconds and 90 seconds, roughly two orders of magnitude faster than eGFRD (see Figure 5 of Chew et al.~\cite{chew2018reaction} and Table \ref{table:updated_benchmark} below). 

%EDIT
In the point particle simulations, the data set (computed before run time) consisted of 3000 linearly spaced distances from $R_{enc}$ to 2*$d_{maxD}$, where $d_{maxD}$ corresponds to the mean square displacement due to diffusion at $T_{maxD}$, the max allowed diffusion time. At each distance, the integrated propensity was computed at 50,000 time points (i.e. variances). There were 40,000 linearly spaced time points from $1*10^{-6}$ to $1*10^{-2}$ where the curvature is often highest, and 10,000 linearly spaced time points from $1*10^{-2}$ to $T_{maxD}$. This 3000 by 50,000 data set was computed in 14.5 minutes in the Go language (golang). 
The propensity function integration error for a given distance value and time duration depends on the number of integration intervals - here 50,000 for the full duration. We used the trapezoid method, whose error at each time point can be upper bounded by $Err(\Delta t) = \frac{\Delta t^{3}}{12N^{2}}*K*c$, where K is the maximum magnitude of the second derivative of the CDF, c is the intrinsic reaction rate, and N is the number of integration intervals over the duration $\Delta t$. These are not likely to be tight upper bounds due to the presence of inflection points in the CDF graph. Our golang integration code, including a method to print error bounds and integrals to text files, can be found in the GitHub repository.
%EDIT

In the finite particle simulations, the data set consisted of 3000 integrated propensity curves at the same distance values, each of which was evaluated at 5500 time points. Computation of the integrated propensity data set required $\sim 1m$. Performing these numerical integrations is possible in Matlab but requires symbolic computation to evaluate the integrands. The result was that the same data set takes on the order of days to compute. In golang, the necessary numeric precision was achieved using its big math package which implements arbitrary-precision arithmetic. 

\begin{figure*}[!]
  \includegraphics[width=\textwidth]{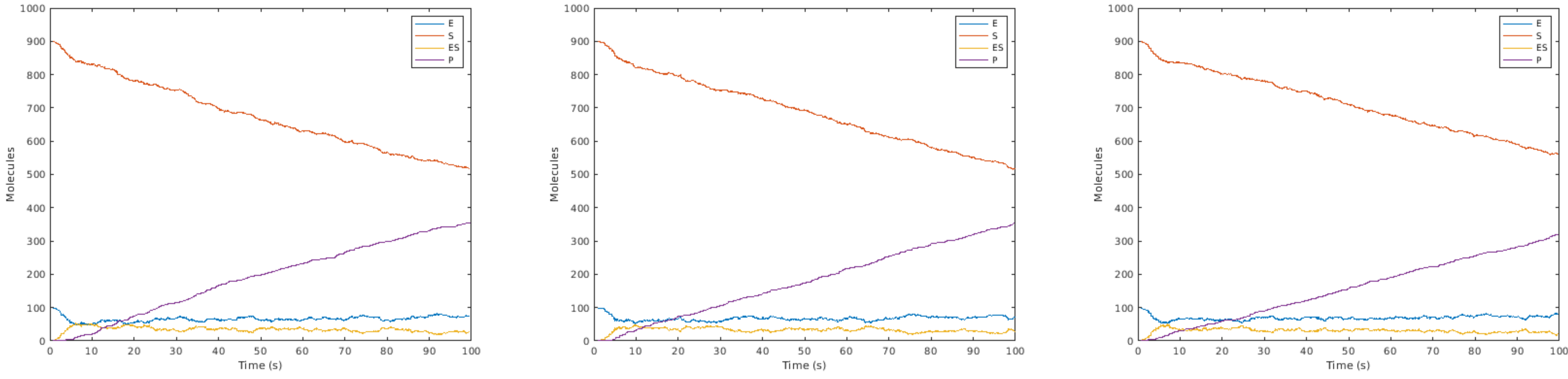}
  \caption{Varying the simulation volume's boundary parameter for the finite particle representation. Time evolution of 1000 molecules in the Michaelis-Menten model with DESSA-CS. Unimolecular rate constant $k_{uni} = 0.1s^{-1}$ (governing $ES \Rightarrow E + S$ and $ES \Rightarrow E + P$), diffusion coefficient $D = 1 \mu m^{2} s^{-1}$, boundary condition parameter $c = 1.2*10^{3}s^{-1}$, and $R_{enc} = 0.01\mu m$.  From left to right: $l_{B}=0.03$ (runtime = 2693s), $l_{B}=0.1$ (runtime = 777s), and $l_{B}=0.3$ (runtime = 157s).}
  \label{fig:MM_finte_exceedBoundarComparisons}
\end{figure*}

In general, both the finite and point particle representations run more efficiently when each molecule is allowed to diffuse farther outside the boundaries before wait times are sampled. We will refer to a cubic simulation volume as having dimension $(L\mu m)^{3}$ and the fraction of the cube length beyond which a particle may diffuse as $l_{B}$.  It is then useful to analyze the behavior of simulations as these vary. The Michaelis-Menten Benchmark requires that $90\mu m^3 = L(1 + l_{B})$. In Figure \ref{fig:MM_finte_exceedBoundarComparisons}, we plot simulation trajectories at multiple values of $l_{B}$ (i.e., $l_{B}=0.03$, $l_{B}=0.1$, and $l_{B}=0.3$) for the finite particle representation.  The plot demonstrates that while the kinetics do not change appreciably with changes in $l_{B}$, the run time does. The respective run times are 2693s, 777s, and 157s. The results are qualitatively similar in the point particle representation case. As the relative distance allotted to the cube's length increases, so does the run time. This results from the fact that as $l_{B}$ decreases, so does the maximum diffusion time of molecules near the boundary, leading to a much higher number of position updates compared with the roughly unchanging number of reaction events. Even though the change in kinetics is minimal across the parameter values examined, the effective association rate does show a weak inverse dependence on $l_{B}$.

\iffalse
An important difference between the point particle and finite particle representations is that in the former case, $Pr(encounter \hspace{1mm}|\hspace{1mm} t)$ does not depend on the intrinsic reaction rate while in the latter case it does. This allowed us to compute integrated encounter probabilities before simulation time when considering point particles. The integrated propensities used during the simulation were obtained from these by multiplication with the intrinsic rate. The implication is that for a single pre-computed data set, simulations can be run with many values of the diffusion rates and intrinsic reaction rates. For models capable of producing large or complex oligomers, it may become necessary to compute such data sets at multiple $R_{enc}$ values. In the finite particle representation, integrated propensity data sets must be pre-computed for each combination, [($D_{a}+D_{b}$),$c$,$R_{enc}$].
\fi

\section{Conclusions}

We have presented a novel event-based method for simulating reaction diffusion systems in continuous space and in the presence of planar or curved boundaries. As in the Gillespie algorithm and related methods, we sample bimolecular reaction waiting times by utilizing propensity functions. However, with the introduction of 3d space, the reaction propensities now depend explicitly on the time reactants diffuse, allowing them to encounter one another. The result is that we integrate the propensity function of each reactant pair in order to determine whether (and when) a reaction is possible in a specified duration. While the method is inspired by ideas from GFRD and eGFRD, our method for sampling reaction locations given the waiting time is, to our knowledge, novel relative to other spatial simulation methods. For point particles, we rely on two assumptions: (1) that reactions must happen in the region both reactants' diffusion spheres overlap and (2) that the probability distributions characterizing the possible distances either reactant has traveled are Gaussian and independent. This implies there are rings of equiprobable points at constant distance from the Gaussian means of the reactants and suggests a method to sample such a ring: first, sample a distance, $r_{A}$ from one reactant, and then sample the angle w.r.t.~the axis connecting the means given $r_{A}$. Each ring is uniquely determined by this distance and angle, and the reaction location can then be selected uniformly at random from on the ring. In the case of molecules with finite size, Green's functions governing radial separation must be used for wait time sampling, however the same ring sampling procedure applies to reaction locations.
We compare our method with its most relevant competitor, eGFRD, on the modified Michaelis-Menten benchmark model described in Chew et al.\cite{chew2018reaction}. The dynamics displayed by eGFRD, Spatiocyte (implementing a microscopic lattice method), and Smoldyn are quantitatively similar.  DESSA-CS shows a substantial improvement over the run time of eGFRD, achieving run times more comparable to the discrete-time alternatives Smoldyn and Spaciocyte MLM.
The method as presented leaves several avenues for extension and improvement in future work.  DESSA-CS is able to achieve its comparatively high run time efficiency by exploiting the fact that, under certain assumptions, wait time sampling can be described by a deterministic part applicable in many circumstances, and a stochastic part specific to each reactant pair. We can therefore perform much of the expensive deterministic computations once, independently of each simulation run. Those assumptions include isotropic diffusion as the primary method of transport, and that reactions between distinct pairs of molecules are described by time-inhomogenous Poisson processes with mean parameter equal to the integrated propensity (this implies exponentially distributed waiting times). These are reasonable assumptions, yet both may be relaxed in future work. Numerically integrating these reaction propensities when considering new reactions at every step of the simulation can lead to the same computations being performed thousands or millions of times. Only the sampling of the exponentially distributed random numbers must be performed for all potential bimolecular reactions. The overall accuracy of the method is dependent on the resolution of the pre-computed integrated propensity curves. In future work, we will investigate more formally the nature of the accuracy/efficiency trade off associated with our handling of reflective boundaries. We will also consider methods for updating our unconstrained integrated reaction propensities to account for the boundaries of the simulation volume. One other avenue for improvement is the choice of programming language for the simulator. Matlab was chosen for its ease of use, as well as its testing and plotting infrastructure. Moving fully to Golang or C would likely lead to substantially greater run time efficiency.

\begin{acknowledgments}
M.T. was supported in part by US National Institutes of Health award T32EB009403. R.S. was supported in part by U.S. National Institutes of Health awards 1R21CA216452 and 1R01HG010589.  We are grateful to Dr. Jim Faeder for helpful advice on this project.
\end{acknowledgments}

\appendix

\section{Evaluating $CDF_{Q}(y)$}
\label{A1}

Following Mathai and Provost \cite{mathai1992quadratic}, we consider the $p$ dimensional Gaussian distributed random variable $X \sim N(\mu,\Sigma)$, $\sum > 0$ and the quadratic form $Q = X^{T}AX$, $A^{T} = A$. $Q$ has the following alternate representation in terms of its eigenvalues:
\begin{equation}
Q = \sum_{j=1}^{p} \lambda_{j}(U_{j} + b_{j})^2 
\end{equation}
Let P be a $p$x$p$ matrix which diagonalizes $\Sigma^{1/2}A\Sigma^{1/2}$, i.e.  $P^{T}\Sigma^{1/2}A\Sigma^{1/2}P = diag(\lambda_{1},...,\lambda_{p})$, and $PP^{T} = I$. Now, $\mathbf{U^{T}} = (U_{1},...,U_{p})$, $\mathbf{U} = P^{T}\Sigma^{-1/2}(X-\mathbf{\mu})$, $\mathbf{b^{T}} = (b_{1},...,b_{p}) = (P^{T}\Sigma^{-1/2}\mathbf{\mu})^{T}$, and the $U_{j}$'s are mutually independent standard normal variables.
We are interested in the distribution function (i.e. CDF of Q) which we here define as $F_{p}(\mathbf{\lambda},\mathbf{b};y)$.
It can be shown (see sections 4.1-4.2) that

\begin{equation}
    F_{p}(\mathbf{\lambda},\mathbf{b};y) = \sum_{k=0}^{\infty} (-1)^{k} z_{k} \frac{y^{p/2 + k}}{\Gamma(p/2 + k + 1)}, \hspace{3mm}
\end{equation}
$0<y<\infty$, with the following recursively defined coefficients.
\begin{equation}
z_{0} = exp(-\frac{1}{2}\sum_{j=1}^{p}b_{j}^{2})\prod_{j=1}^{p}(2\lambda_{j})^{-1/2}
\end{equation}
\begin{equation}
z_{k} = \frac{1}{k}\sum_{r=0}^{k-1}d_{k-r}z_{r}, \hspace{3mm} k\geq1
\end{equation}
\begin{equation}
d_{k} = \frac{1}{2}\sum_{j=1}^{p}(1-kb_{j}^{2})(2\lambda_{j})^{-k}, \hspace{3mm} k\geq1
\end{equation}

%%\nocite{*}

%\input{main.bbl}
\bibliography{Bibliography.bib}% Produces the bibliography via BibTeX.

\begin{thebibliography}{52}%
\makeatletter
\providecommand \@ifxundefined [1]{%
 \@ifx{#1\undefined}
}%
\providecommand \@ifnum [1]{%
 \ifnum #1\expandafter \@firstoftwo
 \else \expandafter \@secondoftwo
 \fi
}%
\providecommand \@ifx [1]{%
 \ifx #1\expandafter \@firstoftwo
 \else \expandafter \@secondoftwo
 \fi
}%
\providecommand \natexlab [1]{#1}%
\providecommand \enquote  [1]{``#1''}%
\providecommand \bibnamefont  [1]{#1}%
\providecommand \bibfnamefont [1]{#1}%
\providecommand \citenamefont [1]{#1}%
\providecommand \href@noop [0]{\@secondoftwo}%
\providecommand \href [0]{\begingroup \@sanitize@url \@href}%
\providecommand \@href[1]{\@@startlink{#1}\@@href}%
\providecommand \@@href[1]{\endgroup#1\@@endlink}%
\providecommand \@sanitize@url [0]{\catcode `\\12\catcode `\$12\catcode
  `\&12\catcode `\#12\catcode `\^12\catcode `\_12\catcode `\%12\relax}%
\providecommand \@@startlink[1]{}%
\providecommand \@@endlink[0]{}%
\providecommand \url  [0]{\begingroup\@sanitize@url \@url }%
\providecommand \@url [1]{\endgroup\@href {#1}{\urlprefix }}%
\providecommand \urlprefix  [0]{URL }%
\providecommand \Eprint [0]{\href }%
\providecommand \doibase [0]{https://doi.org/}%
\providecommand \selectlanguage [0]{\@gobble}%
\providecommand \bibinfo  [0]{\@secondoftwo}%
\providecommand \bibfield  [0]{\@secondoftwo}%
\providecommand \translation [1]{[#1]}%
\providecommand \BibitemOpen [0]{}%
\providecommand \bibitemStop [0]{}%
\providecommand \bibitemNoStop [0]{.\EOS\space}%
\providecommand \EOS [0]{\spacefactor3000\relax}%
\providecommand \BibitemShut  [1]{\csname bibitem#1\endcsname}%
\let\auto@bib@innerbib\@empty
%</preamble>
\bibitem [{\citenamefont {Kaya}\ \emph {et~al.}(2018)\citenamefont {Kaya},
  \citenamefont {Cheng}, \citenamefont {Block}, \citenamefont {Bartol},
  \citenamefont {Sejnowski}, \citenamefont {Sorkin}, \citenamefont {Faeder},\
  and\ \citenamefont {Bahar}}]{kaya2018heterogeneities}%
  \BibitemOpen
  \bibfield  {author} {\bibinfo {author} {\bibfnamefont {C.}~\bibnamefont
  {Kaya}}, \bibinfo {author} {\bibfnamefont {M.~H.}\ \bibnamefont {Cheng}},
  \bibinfo {author} {\bibfnamefont {E.~R.}\ \bibnamefont {Block}}, \bibinfo
  {author} {\bibfnamefont {T.~M.}\ \bibnamefont {Bartol}}, \bibinfo {author}
  {\bibfnamefont {T.~J.}\ \bibnamefont {Sejnowski}}, \bibinfo {author}
  {\bibfnamefont {A.}~\bibnamefont {Sorkin}}, \bibinfo {author} {\bibfnamefont
  {J.~R.}\ \bibnamefont {Faeder}},\ and\ \bibinfo {author} {\bibfnamefont
  {I.}~\bibnamefont {Bahar}},\ }\bibfield  {title} {\bibinfo {title}
  {Heterogeneities in axonal structure and transporter distribution lower
  dopamine reuptake efficiency},\ }\href@noop {} {\bibfield  {journal}
  {\bibinfo  {journal} {eNeuro}\ }\textbf {\bibinfo {volume} {5}} (\bibinfo
  {year} {2018})}\BibitemShut {NoStop}%
\bibitem [{\citenamefont {Thomas}\ and\ \citenamefont
  {Schwartz}(2017)}]{thomas2017quantitative}%
  \BibitemOpen
  \bibfield  {author} {\bibinfo {author} {\bibfnamefont {M.}~\bibnamefont
  {Thomas}}\ and\ \bibinfo {author} {\bibfnamefont {R.}~\bibnamefont
  {Schwartz}},\ }\bibfield  {title} {\bibinfo {title} {Quantitative
  computational models of molecular self-assembly in systems biology},\
  }\href@noop {} {\bibfield  {journal} {\bibinfo  {journal} {Physical biology}\
  }\textbf {\bibinfo {volume} {14}},\ \bibinfo {pages} {035003} (\bibinfo
  {year} {2017})}\BibitemShut {NoStop}%
\bibitem [{\citenamefont {Chevreuil}\ \emph {et~al.}(2018)\citenamefont
  {Chevreuil}, \citenamefont {Law-Hine}, \citenamefont {Chen}, \citenamefont
  {Bressanelli}, \citenamefont {Combet}, \citenamefont {Constantin},
  \citenamefont {Degrouard}, \citenamefont {M{\"o}ller}, \citenamefont
  {Zeghal},\ and\ \citenamefont {Tresset}}]{chevreuil2018nonequilibrium}%
  \BibitemOpen
  \bibfield  {author} {\bibinfo {author} {\bibfnamefont {M.}~\bibnamefont
  {Chevreuil}}, \bibinfo {author} {\bibfnamefont {D.}~\bibnamefont {Law-Hine}},
  \bibinfo {author} {\bibfnamefont {J.}~\bibnamefont {Chen}}, \bibinfo {author}
  {\bibfnamefont {S.}~\bibnamefont {Bressanelli}}, \bibinfo {author}
  {\bibfnamefont {S.}~\bibnamefont {Combet}}, \bibinfo {author} {\bibfnamefont
  {D.}~\bibnamefont {Constantin}}, \bibinfo {author} {\bibfnamefont
  {J.}~\bibnamefont {Degrouard}}, \bibinfo {author} {\bibfnamefont
  {J.}~\bibnamefont {M{\"o}ller}}, \bibinfo {author} {\bibfnamefont
  {M.}~\bibnamefont {Zeghal}},\ and\ \bibinfo {author} {\bibfnamefont
  {G.}~\bibnamefont {Tresset}},\ }\bibfield  {title} {\bibinfo {title}
  {Nonequilibrium self-assembly dynamics of icosahedral viral capsids packaging
  genome or polyelectrolyte},\ }\href@noop {} {\bibfield  {journal} {\bibinfo
  {journal} {Nature communications}\ }\textbf {\bibinfo {volume} {9}},\
  \bibinfo {pages} {3071} (\bibinfo {year} {2018})}\BibitemShut {NoStop}%
\bibitem [{\citenamefont {Lopez-Fontal}\ \emph {et~al.}(2018)\citenamefont
  {Lopez-Fontal}, \citenamefont {Grochmal}, \citenamefont {Foran},
  \citenamefont {Milanesi},\ and\ \citenamefont {Tomas}}]{lopez2018ship}%
  \BibitemOpen
  \bibfield  {author} {\bibinfo {author} {\bibfnamefont {E.}~\bibnamefont
  {Lopez-Fontal}}, \bibinfo {author} {\bibfnamefont {A.}~\bibnamefont
  {Grochmal}}, \bibinfo {author} {\bibfnamefont {T.}~\bibnamefont {Foran}},
  \bibinfo {author} {\bibfnamefont {L.}~\bibnamefont {Milanesi}},\ and\
  \bibinfo {author} {\bibfnamefont {S.}~\bibnamefont {Tomas}},\ }\bibfield
  {title} {\bibinfo {title} {Ship in a bottle: confinement-promoted
  self-assembly},\ }\href@noop {} {\bibfield  {journal} {\bibinfo  {journal}
  {Chemical science}\ }\textbf {\bibinfo {volume} {9}},\ \bibinfo {pages}
  {1760} (\bibinfo {year} {2018})}\BibitemShut {NoStop}%
\bibitem [{\citenamefont {Wang}\ \emph {et~al.}(2018)\citenamefont {Wang},
  \citenamefont {Hermes}, \citenamefont {Kotni}, \citenamefont {Wu},
  \citenamefont {Tasios}, \citenamefont {Liu}, \citenamefont {De~Nijs},
  \citenamefont {Van Der~Wee}, \citenamefont {Murray}, \citenamefont {Dijkstra}
  \emph {et~al.}}]{wang2018interplay}%
  \BibitemOpen
  \bibfield  {author} {\bibinfo {author} {\bibfnamefont {D.}~\bibnamefont
  {Wang}}, \bibinfo {author} {\bibfnamefont {M.}~\bibnamefont {Hermes}},
  \bibinfo {author} {\bibfnamefont {R.}~\bibnamefont {Kotni}}, \bibinfo
  {author} {\bibfnamefont {Y.}~\bibnamefont {Wu}}, \bibinfo {author}
  {\bibfnamefont {N.}~\bibnamefont {Tasios}}, \bibinfo {author} {\bibfnamefont
  {Y.}~\bibnamefont {Liu}}, \bibinfo {author} {\bibfnamefont {B.}~\bibnamefont
  {De~Nijs}}, \bibinfo {author} {\bibfnamefont {E.~B.}\ \bibnamefont {Van
  Der~Wee}}, \bibinfo {author} {\bibfnamefont {C.~B.}\ \bibnamefont {Murray}},
  \bibinfo {author} {\bibfnamefont {M.}~\bibnamefont {Dijkstra}}, \emph
  {et~al.},\ }\bibfield  {title} {\bibinfo {title} {Interplay between spherical
  confinement and particle shape on the self-assembly of rounded cubes},\
  }\href@noop {} {\bibfield  {journal} {\bibinfo  {journal} {Nature
  communications}\ }\textbf {\bibinfo {volume} {9}},\ \bibinfo {pages} {2228}
  (\bibinfo {year} {2018})}\BibitemShut {NoStop}%
\bibitem [{\citenamefont {Junker}\ \emph {et~al.}(2019)\citenamefont {Junker},
  \citenamefont {Vaghefikia}, \citenamefont {Albarghash}, \citenamefont
  {H{\"o}fig}, \citenamefont {Kempe}, \citenamefont {Walter}, \citenamefont
  {Otten}, \citenamefont {Pohl}, \citenamefont {Katranidis}, \citenamefont
  {Wiegand} \emph {et~al.}}]{junker2019impact}%
  \BibitemOpen
  \bibfield  {author} {\bibinfo {author} {\bibfnamefont {N.~O.}\ \bibnamefont
  {Junker}}, \bibinfo {author} {\bibfnamefont {F.}~\bibnamefont {Vaghefikia}},
  \bibinfo {author} {\bibfnamefont {A.}~\bibnamefont {Albarghash}}, \bibinfo
  {author} {\bibfnamefont {H.}~\bibnamefont {H{\"o}fig}}, \bibinfo {author}
  {\bibfnamefont {D.}~\bibnamefont {Kempe}}, \bibinfo {author} {\bibfnamefont
  {J.}~\bibnamefont {Walter}}, \bibinfo {author} {\bibfnamefont
  {J.}~\bibnamefont {Otten}}, \bibinfo {author} {\bibfnamefont
  {M.}~\bibnamefont {Pohl}}, \bibinfo {author} {\bibfnamefont {A.}~\bibnamefont
  {Katranidis}}, \bibinfo {author} {\bibfnamefont {S.}~\bibnamefont {Wiegand}},
  \emph {et~al.},\ }\bibfield  {title} {\bibinfo {title} {The impact of
  molecular crowding on translational mobility and conformational properties of
  biological macromolecules},\ }\href@noop {} {\bibfield  {journal} {\bibinfo
  {journal} {The Journal of Physical Chemistry B}\ } (\bibinfo {year}
  {2019})}\BibitemShut {NoStop}%
\bibitem [{\citenamefont {Smith}\ \emph {et~al.}(2014)\citenamefont {Smith},
  \citenamefont {Xie}, \citenamefont {Lee},\ and\ \citenamefont
  {Schwartz}}]{smith2014applying}%
  \BibitemOpen
  \bibfield  {author} {\bibinfo {author} {\bibfnamefont {G.~R.}\ \bibnamefont
  {Smith}}, \bibinfo {author} {\bibfnamefont {L.}~\bibnamefont {Xie}}, \bibinfo
  {author} {\bibfnamefont {B.}~\bibnamefont {Lee}},\ and\ \bibinfo {author}
  {\bibfnamefont {R.}~\bibnamefont {Schwartz}},\ }\bibfield  {title} {\bibinfo
  {title} {Applying molecular crowding models to simulations of virus capsid
  assembly in vitro},\ }\href@noop {} {\bibfield  {journal} {\bibinfo
  {journal} {Biophysical journal}\ }\textbf {\bibinfo {volume} {106}},\
  \bibinfo {pages} {310} (\bibinfo {year} {2014})}\BibitemShut {NoStop}%
\bibitem [{\citenamefont {Van~Treeck}\ \emph {et~al.}(2018)\citenamefont
  {Van~Treeck}, \citenamefont {Protter}, \citenamefont {Matheny}, \citenamefont
  {Khong}, \citenamefont {Link},\ and\ \citenamefont {Parker}}]{van2018rna}%
  \BibitemOpen
  \bibfield  {author} {\bibinfo {author} {\bibfnamefont {B.}~\bibnamefont
  {Van~Treeck}}, \bibinfo {author} {\bibfnamefont {D.~S.}\ \bibnamefont
  {Protter}}, \bibinfo {author} {\bibfnamefont {T.}~\bibnamefont {Matheny}},
  \bibinfo {author} {\bibfnamefont {A.}~\bibnamefont {Khong}}, \bibinfo
  {author} {\bibfnamefont {C.~D.}\ \bibnamefont {Link}},\ and\ \bibinfo
  {author} {\bibfnamefont {R.}~\bibnamefont {Parker}},\ }\bibfield  {title}
  {\bibinfo {title} {Rna self-assembly contributes to stress granule formation
  and defining the stress granule transcriptome},\ }\href@noop {} {\bibfield
  {journal} {\bibinfo  {journal} {Proceedings of the National Academy of
  Sciences}\ }\textbf {\bibinfo {volume} {115}},\ \bibinfo {pages} {2734}
  (\bibinfo {year} {2018})}\BibitemShut {NoStop}%
\bibitem [{\citenamefont {Gillespie}(1992)}]{gillespie1992rigorous}%
  \BibitemOpen
  \bibfield  {author} {\bibinfo {author} {\bibfnamefont {D.~T.}\ \bibnamefont
  {Gillespie}},\ }\bibfield  {title} {\bibinfo {title} {A rigorous derivation
  of the chemical master equation},\ }\href@noop {} {\bibfield  {journal}
  {\bibinfo  {journal} {Physica A: Statistical Mechanics and its Applications}\
  }\textbf {\bibinfo {volume} {188}},\ \bibinfo {pages} {404} (\bibinfo {year}
  {1992})}\BibitemShut {NoStop}%
\bibitem [{\citenamefont {Gillespie}(2009)}]{gillespie2009diffusional}%
  \BibitemOpen
  \bibfield  {author} {\bibinfo {author} {\bibfnamefont {D.~T.}\ \bibnamefont
  {Gillespie}},\ }\bibfield  {title} {\bibinfo {title} {A diffusional
  bimolecular propensity function},\ }\href@noop {} {\bibfield  {journal}
  {\bibinfo  {journal} {The Journal of chemical physics}\ }\textbf {\bibinfo
  {volume} {131}},\ \bibinfo {pages} {164109} (\bibinfo {year}
  {2009})}\BibitemShut {NoStop}%
\bibitem [{\citenamefont {Gillespie}\ \emph {et~al.}(2013)\citenamefont
  {Gillespie}, \citenamefont {Hellander},\ and\ \citenamefont
  {Petzold}}]{gillespie2013perspective}%
  \BibitemOpen
  \bibfield  {author} {\bibinfo {author} {\bibfnamefont {D.~T.}\ \bibnamefont
  {Gillespie}}, \bibinfo {author} {\bibfnamefont {A.}~\bibnamefont
  {Hellander}},\ and\ \bibinfo {author} {\bibfnamefont {L.~R.}\ \bibnamefont
  {Petzold}},\ }\bibfield  {title} {\bibinfo {title} {Perspective: Stochastic
  algorithms for chemical kinetics},\ }\href@noop {} {\bibfield  {journal}
  {\bibinfo  {journal} {The Journal of chemical physics}\ }\textbf {\bibinfo
  {volume} {138}},\ \bibinfo {pages} {05B201\_1} (\bibinfo {year}
  {2013})}\BibitemShut {NoStop}%
\bibitem [{\citenamefont {Warne}\ \emph {et~al.}(2019)\citenamefont {Warne},
  \citenamefont {Baker},\ and\ \citenamefont {Simpson}}]{warne2019simulation}%
  \BibitemOpen
  \bibfield  {author} {\bibinfo {author} {\bibfnamefont {D.~J.}\ \bibnamefont
  {Warne}}, \bibinfo {author} {\bibfnamefont {R.~E.}\ \bibnamefont {Baker}},\
  and\ \bibinfo {author} {\bibfnamefont {M.~J.}\ \bibnamefont {Simpson}},\
  }\bibfield  {title} {\bibinfo {title} {Simulation and inference algorithms
  for stochastic biochemical reaction networks: from basic concepts to
  state-of-the-art},\ }\href@noop {} {\bibfield  {journal} {\bibinfo  {journal}
  {Journal of the Royal Society Interface}\ }\textbf {\bibinfo {volume} {16}},\
  \bibinfo {pages} {20180943} (\bibinfo {year} {2019})}\BibitemShut {NoStop}%
\bibitem [{\citenamefont {Nag}\ \emph {et~al.}(2009)\citenamefont {Nag},
  \citenamefont {Monine}, \citenamefont {Faeder},\ and\ \citenamefont
  {Goldstein}}]{nag2009aggregation}%
  \BibitemOpen
  \bibfield  {author} {\bibinfo {author} {\bibfnamefont {A.}~\bibnamefont
  {Nag}}, \bibinfo {author} {\bibfnamefont {M.~I.}\ \bibnamefont {Monine}},
  \bibinfo {author} {\bibfnamefont {J.~R.}\ \bibnamefont {Faeder}},\ and\
  \bibinfo {author} {\bibfnamefont {B.}~\bibnamefont {Goldstein}},\ }\bibfield
  {title} {\bibinfo {title} {Aggregation of membrane proteins by cytosolic
  cross-linkers: theory and simulation of the lat-grb2-sos1 system},\
  }\href@noop {} {\bibfield  {journal} {\bibinfo  {journal} {Biophysical
  journal}\ }\textbf {\bibinfo {volume} {96}},\ \bibinfo {pages} {2604}
  (\bibinfo {year} {2009})}\BibitemShut {NoStop}%
\bibitem [{\citenamefont {Gillespie}(2001)}]{gillespie2001approximate}%
  \BibitemOpen
  \bibfield  {author} {\bibinfo {author} {\bibfnamefont {D.~T.}\ \bibnamefont
  {Gillespie}},\ }\bibfield  {title} {\bibinfo {title} {Approximate accelerated
  stochastic simulation of chemically reacting systems},\ }\href@noop {}
  {\bibfield  {journal} {\bibinfo  {journal} {The Journal of Chemical Physics}\
  }\textbf {\bibinfo {volume} {115}},\ \bibinfo {pages} {1716} (\bibinfo {year}
  {2001})}\BibitemShut {NoStop}%
\bibitem [{\citenamefont {Rathinam}\ \emph {et~al.}(2003)\citenamefont
  {Rathinam}, \citenamefont {Petzold}, \citenamefont {Cao},\ and\ \citenamefont
  {Gillespie}}]{rathinam2003stiffness}%
  \BibitemOpen
  \bibfield  {author} {\bibinfo {author} {\bibfnamefont {M.}~\bibnamefont
  {Rathinam}}, \bibinfo {author} {\bibfnamefont {L.~R.}\ \bibnamefont
  {Petzold}}, \bibinfo {author} {\bibfnamefont {Y.}~\bibnamefont {Cao}},\ and\
  \bibinfo {author} {\bibfnamefont {D.~T.}\ \bibnamefont {Gillespie}},\
  }\bibfield  {title} {\bibinfo {title} {Stiffness in stochastic chemically
  reacting systems: The implicit tau-leaping method},\ }\href@noop {}
  {\bibfield  {journal} {\bibinfo  {journal} {The Journal of Chemical Physics}\
  }\textbf {\bibinfo {volume} {119}},\ \bibinfo {pages} {12784} (\bibinfo
  {year} {2003})}\BibitemShut {NoStop}%
\bibitem [{\citenamefont {Cao}\ \emph {et~al.}(2005)\citenamefont {Cao},
  \citenamefont {Gillespie},\ and\ \citenamefont {Petzold}}]{cao2005slow}%
  \BibitemOpen
  \bibfield  {author} {\bibinfo {author} {\bibfnamefont {Y.}~\bibnamefont
  {Cao}}, \bibinfo {author} {\bibfnamefont {D.~T.}\ \bibnamefont {Gillespie}},\
  and\ \bibinfo {author} {\bibfnamefont {L.~R.}\ \bibnamefont {Petzold}},\
  }\bibfield  {title} {\bibinfo {title} {The slow-scale stochastic simulation
  algorithm},\ }\href@noop {} {\bibfield  {journal} {\bibinfo  {journal} {The
  Journal of chemical physics}\ }\textbf {\bibinfo {volume} {122}},\ \bibinfo
  {pages} {014116} (\bibinfo {year} {2005})}\BibitemShut {NoStop}%
\bibitem [{\citenamefont {Jamalyaria}\ \emph {et~al.}(2005)\citenamefont
  {Jamalyaria}, \citenamefont {Rohlfs},\ and\ \citenamefont
  {Schwartz}}]{jamalyaria2005queue}%
  \BibitemOpen
  \bibfield  {author} {\bibinfo {author} {\bibfnamefont {F.}~\bibnamefont
  {Jamalyaria}}, \bibinfo {author} {\bibfnamefont {R.}~\bibnamefont {Rohlfs}},\
  and\ \bibinfo {author} {\bibfnamefont {R.}~\bibnamefont {Schwartz}},\
  }\bibfield  {title} {\bibinfo {title} {Queue-based method for efficient
  simulation of biological self-assembly systems},\ }\href@noop {} {\bibfield
  {journal} {\bibinfo  {journal} {Journal of Computational Physics}\ }\textbf
  {\bibinfo {volume} {204}},\ \bibinfo {pages} {100} (\bibinfo {year}
  {2005})}\BibitemShut {NoStop}%
\bibitem [{\citenamefont {Misra}\ and\ \citenamefont
  {Schwartz}(2008)}]{misra2008efficient}%
  \BibitemOpen
  \bibfield  {author} {\bibinfo {author} {\bibfnamefont {N.}~\bibnamefont
  {Misra}}\ and\ \bibinfo {author} {\bibfnamefont {R.}~\bibnamefont
  {Schwartz}},\ }\bibfield  {title} {\bibinfo {title} {Efficient stochastic
  sampling of first-passage times with applications to self-assembly
  simulations},\ }\href@noop {} {\bibfield  {journal} {\bibinfo  {journal} {The
  Journal of chemical physics}\ }\textbf {\bibinfo {volume} {129}},\ \bibinfo
  {pages} {204109} (\bibinfo {year} {2008})}\BibitemShut {NoStop}%
\bibitem [{\citenamefont {Anderson}(2008)}]{anderson2008incorporating}%
  \BibitemOpen
  \bibfield  {author} {\bibinfo {author} {\bibfnamefont {D.~F.}\ \bibnamefont
  {Anderson}},\ }\bibfield  {title} {\bibinfo {title} {Incorporating postleap
  checks in tau-leaping},\ }\href@noop {} {\bibfield  {journal} {\bibinfo
  {journal} {The Journal of chemical physics}\ }\textbf {\bibinfo {volume}
  {128}},\ \bibinfo {pages} {054103} (\bibinfo {year} {2008})}\BibitemShut
  {NoStop}%
\bibitem [{\citenamefont {Sneddon}\ \emph {et~al.}(2011)\citenamefont
  {Sneddon}, \citenamefont {Faeder},\ and\ \citenamefont
  {Emonet}}]{sneddon2011efficient}%
  \BibitemOpen
  \bibfield  {author} {\bibinfo {author} {\bibfnamefont {M.~W.}\ \bibnamefont
  {Sneddon}}, \bibinfo {author} {\bibfnamefont {J.~R.}\ \bibnamefont
  {Faeder}},\ and\ \bibinfo {author} {\bibfnamefont {T.}~\bibnamefont
  {Emonet}},\ }\bibfield  {title} {\bibinfo {title} {Efficient modeling,
  simulation and coarse-graining of biological complexity with nfsim},\
  }\href@noop {} {\bibfield  {journal} {\bibinfo  {journal} {Nature methods}\
  }\textbf {\bibinfo {volume} {8}},\ \bibinfo {pages} {177} (\bibinfo {year}
  {2011})}\BibitemShut {NoStop}%
\bibitem [{\citenamefont {Donovan}\ \emph {et~al.}(2013)\citenamefont
  {Donovan}, \citenamefont {Sedgewick}, \citenamefont {Faeder},\ and\
  \citenamefont {Zuckerman}}]{donovan2013efficient}%
  \BibitemOpen
  \bibfield  {author} {\bibinfo {author} {\bibfnamefont {R.~M.}\ \bibnamefont
  {Donovan}}, \bibinfo {author} {\bibfnamefont {A.~J.}\ \bibnamefont
  {Sedgewick}}, \bibinfo {author} {\bibfnamefont {J.~R.}\ \bibnamefont
  {Faeder}},\ and\ \bibinfo {author} {\bibfnamefont {D.~M.}\ \bibnamefont
  {Zuckerman}},\ }\bibfield  {title} {\bibinfo {title} {Efficient stochastic
  simulation of chemical kinetics networks using a weighted ensemble of
  trajectories},\ }\href@noop {} {\bibfield  {journal} {\bibinfo  {journal}
  {The Journal of chemical physics}\ }\textbf {\bibinfo {volume} {139}},\
  \bibinfo {pages} {09B642\_1} (\bibinfo {year} {2013})}\BibitemShut {NoStop}%
\bibitem [{\citenamefont {Lin}\ \emph {et~al.}(2019)\citenamefont {Lin},
  \citenamefont {Feng},\ and\ \citenamefont {Hlavacek}}]{lin2019scaling}%
  \BibitemOpen
  \bibfield  {author} {\bibinfo {author} {\bibfnamefont {Y.~T.}\ \bibnamefont
  {Lin}}, \bibinfo {author} {\bibfnamefont {S.}~\bibnamefont {Feng}},\ and\
  \bibinfo {author} {\bibfnamefont {W.~S.}\ \bibnamefont {Hlavacek}},\
  }\bibfield  {title} {\bibinfo {title} {Scaling methods for accelerating
  kinetic monte carlo simulations of chemical reaction networks},\ }\href@noop
  {} {\bibfield  {journal} {\bibinfo  {journal} {The Journal of Chemical
  Physics}\ }\textbf {\bibinfo {volume} {150}},\ \bibinfo {pages} {244101}
  (\bibinfo {year} {2019})}\BibitemShut {NoStop}%
\bibitem [{\citenamefont {Baras}\ and\ \citenamefont
  {Mansour}(1996)}]{baras1996reaction}%
  \BibitemOpen
  \bibfield  {author} {\bibinfo {author} {\bibfnamefont {F.}~\bibnamefont
  {Baras}}\ and\ \bibinfo {author} {\bibfnamefont {M.~M.}\ \bibnamefont
  {Mansour}},\ }\bibfield  {title} {\bibinfo {title} {Reaction-diffusion master
  equation: A comparison with microscopic simulations},\ }\href@noop {}
  {\bibfield  {journal} {\bibinfo  {journal} {Physical Review E}\ }\textbf
  {\bibinfo {volume} {54}},\ \bibinfo {pages} {6139} (\bibinfo {year}
  {1996})}\BibitemShut {NoStop}%
\bibitem [{\citenamefont {Isaacson}(2009)}]{isaacson2009reaction}%
  \BibitemOpen
  \bibfield  {author} {\bibinfo {author} {\bibfnamefont {S.~A.}\ \bibnamefont
  {Isaacson}},\ }\bibfield  {title} {\bibinfo {title} {The reaction-diffusion
  master equation as an asymptotic approximation of diffusion to a small
  target},\ }\href@noop {} {\bibfield  {journal} {\bibinfo  {journal} {SIAM
  Journal on Applied Mathematics}\ }\textbf {\bibinfo {volume} {70}},\ \bibinfo
  {pages} {77} (\bibinfo {year} {2009})}\BibitemShut {NoStop}%
\bibitem [{\citenamefont {Smith}\ and\ \citenamefont
  {Grima}(2016)}]{smith2016breakdown}%
  \BibitemOpen
  \bibfield  {author} {\bibinfo {author} {\bibfnamefont {S.}~\bibnamefont
  {Smith}}\ and\ \bibinfo {author} {\bibfnamefont {R.}~\bibnamefont {Grima}},\
  }\bibfield  {title} {\bibinfo {title} {Breakdown of the reaction-diffusion
  master equation with nonelementary rates},\ }\href@noop {} {\bibfield
  {journal} {\bibinfo  {journal} {Physical Review E}\ }\textbf {\bibinfo
  {volume} {93}},\ \bibinfo {pages} {052135} (\bibinfo {year}
  {2016})}\BibitemShut {NoStop}%
\bibitem [{\citenamefont {Schwartz}\ \emph {et~al.}(1998)\citenamefont
  {Schwartz}, \citenamefont {Shor}, \citenamefont {Prevelige~Jr},\ and\
  \citenamefont {Berger}}]{schwartz1998local}%
  \BibitemOpen
  \bibfield  {author} {\bibinfo {author} {\bibfnamefont {R.}~\bibnamefont
  {Schwartz}}, \bibinfo {author} {\bibfnamefont {P.~W.}\ \bibnamefont {Shor}},
  \bibinfo {author} {\bibfnamefont {P.~E.}\ \bibnamefont {Prevelige~Jr}},\ and\
  \bibinfo {author} {\bibfnamefont {B.}~\bibnamefont {Berger}},\ }\bibfield
  {title} {\bibinfo {title} {Local rules simulation of the kinetics of virus
  capsid self-assembly},\ }\href@noop {} {\bibfield  {journal} {\bibinfo
  {journal} {Biophysical journal}\ }\textbf {\bibinfo {volume} {75}},\ \bibinfo
  {pages} {2626} (\bibinfo {year} {1998})}\BibitemShut {NoStop}%
\bibitem [{\citenamefont {Bourov}\ and\ \citenamefont
  {Bhattacharya}(2003)}]{bourov2003role}%
  \BibitemOpen
  \bibfield  {author} {\bibinfo {author} {\bibfnamefont {G.~K.}\ \bibnamefont
  {Bourov}}\ and\ \bibinfo {author} {\bibfnamefont {A.}~\bibnamefont
  {Bhattacharya}},\ }\bibfield  {title} {\bibinfo {title} {The role of
  geometric constraints in amphiphilic self-assembly: A brownian dynamics
  study},\ }\href@noop {} {\bibfield  {journal} {\bibinfo  {journal} {The
  Journal of chemical physics}\ }\textbf {\bibinfo {volume} {119}},\ \bibinfo
  {pages} {9219} (\bibinfo {year} {2003})}\BibitemShut {NoStop}%
\bibitem [{\citenamefont {Hagan}\ and\ \citenamefont
  {Chandler}(2006)}]{hagan2006dynamic}%
  \BibitemOpen
  \bibfield  {author} {\bibinfo {author} {\bibfnamefont {M.~F.}\ \bibnamefont
  {Hagan}}\ and\ \bibinfo {author} {\bibfnamefont {D.}~\bibnamefont
  {Chandler}},\ }\bibfield  {title} {\bibinfo {title} {Dynamic pathways for
  viral capsid assembly},\ }\href@noop {} {\bibfield  {journal} {\bibinfo
  {journal} {Biophysical journal}\ }\textbf {\bibinfo {volume} {91}},\ \bibinfo
  {pages} {42} (\bibinfo {year} {2006})}\BibitemShut {NoStop}%
\bibitem [{\citenamefont {Kerr}\ \emph {et~al.}(2008)\citenamefont {Kerr},
  \citenamefont {Bartol}, \citenamefont {Kaminsky}, \citenamefont {Dittrich},
  \citenamefont {Chang}, \citenamefont {Baden}, \citenamefont {Sejnowski},\
  and\ \citenamefont {Stiles}}]{kerr2008fast}%
  \BibitemOpen
  \bibfield  {author} {\bibinfo {author} {\bibfnamefont {R.~A.}\ \bibnamefont
  {Kerr}}, \bibinfo {author} {\bibfnamefont {T.~M.}\ \bibnamefont {Bartol}},
  \bibinfo {author} {\bibfnamefont {B.}~\bibnamefont {Kaminsky}}, \bibinfo
  {author} {\bibfnamefont {M.}~\bibnamefont {Dittrich}}, \bibinfo {author}
  {\bibfnamefont {J.-C.~J.}\ \bibnamefont {Chang}}, \bibinfo {author}
  {\bibfnamefont {S.~B.}\ \bibnamefont {Baden}}, \bibinfo {author}
  {\bibfnamefont {T.~J.}\ \bibnamefont {Sejnowski}},\ and\ \bibinfo {author}
  {\bibfnamefont {J.~R.}\ \bibnamefont {Stiles}},\ }\bibfield  {title}
  {\bibinfo {title} {Fast monte carlo simulation methods for biological
  reaction-diffusion systems in solution and on surfaces},\ }\href@noop {}
  {\bibfield  {journal} {\bibinfo  {journal} {SIAM journal on scientific
  computing}\ }\textbf {\bibinfo {volume} {30}},\ \bibinfo {pages} {3126}
  (\bibinfo {year} {2008})}\BibitemShut {NoStop}%
\bibitem [{\citenamefont {Castle}\ and\ \citenamefont
  {Odde}(2013)}]{castle2013brownian}%
  \BibitemOpen
  \bibfield  {author} {\bibinfo {author} {\bibfnamefont {B.~T.}\ \bibnamefont
  {Castle}}\ and\ \bibinfo {author} {\bibfnamefont {D.~J.}\ \bibnamefont
  {Odde}},\ }\bibfield  {title} {\bibinfo {title} {Brownian dynamics of subunit
  addition-loss kinetics and thermodynamics in linear polymer self-assembly},\
  }\href@noop {} {\bibfield  {journal} {\bibinfo  {journal} {Biophysical
  journal}\ }\textbf {\bibinfo {volume} {105}},\ \bibinfo {pages} {2528}
  (\bibinfo {year} {2013})}\BibitemShut {NoStop}%
\bibitem [{\citenamefont {Castro-Villarreal}\ \emph {et~al.}(2014)\citenamefont
  {Castro-Villarreal}, \citenamefont {Villada-Balbuena}, \citenamefont
  {M{\'e}ndez-Alcaraz}, \citenamefont {Casta{\~n}eda-Priego},\ and\
  \citenamefont {Estrada-Jim{\'e}nez}}]{castro2014brownian}%
  \BibitemOpen
  \bibfield  {author} {\bibinfo {author} {\bibfnamefont {P.}~\bibnamefont
  {Castro-Villarreal}}, \bibinfo {author} {\bibfnamefont {A.}~\bibnamefont
  {Villada-Balbuena}}, \bibinfo {author} {\bibfnamefont {J.~M.}\ \bibnamefont
  {M{\'e}ndez-Alcaraz}}, \bibinfo {author} {\bibfnamefont {R.}~\bibnamefont
  {Casta{\~n}eda-Priego}},\ and\ \bibinfo {author} {\bibfnamefont
  {S.}~\bibnamefont {Estrada-Jim{\'e}nez}},\ }\bibfield  {title} {\bibinfo
  {title} {A brownian dynamics algorithm for colloids in curved manifolds},\
  }\href@noop {} {\bibfield  {journal} {\bibinfo  {journal} {The Journal of
  chemical physics}\ }\textbf {\bibinfo {volume} {140}},\ \bibinfo {pages}
  {214115} (\bibinfo {year} {2014})}\BibitemShut {NoStop}%
\bibitem [{\citenamefont {Bachmann}\ \emph {et~al.}(2016)\citenamefont
  {Bachmann}, \citenamefont {Petitzon},\ and\ \citenamefont
  {Mognetti}}]{bachmann2016bond}%
  \BibitemOpen
  \bibfield  {author} {\bibinfo {author} {\bibfnamefont {S.~J.}\ \bibnamefont
  {Bachmann}}, \bibinfo {author} {\bibfnamefont {M.}~\bibnamefont {Petitzon}},\
  and\ \bibinfo {author} {\bibfnamefont {B.~M.}\ \bibnamefont {Mognetti}},\
  }\bibfield  {title} {\bibinfo {title} {Bond formation kinetics affects
  self-assembly directed by ligand--receptor interactions},\ }\href@noop {}
  {\bibfield  {journal} {\bibinfo  {journal} {Soft matter}\ }\textbf {\bibinfo
  {volume} {12}},\ \bibinfo {pages} {9585} (\bibinfo {year}
  {2016})}\BibitemShut {NoStop}%
\bibitem [{\citenamefont {Donev}\ \emph {et~al.}(2018)\citenamefont {Donev},
  \citenamefont {Yang},\ and\ \citenamefont {Kim}}]{donev2018efficient}%
  \BibitemOpen
  \bibfield  {author} {\bibinfo {author} {\bibfnamefont {A.}~\bibnamefont
  {Donev}}, \bibinfo {author} {\bibfnamefont {C.-Y.}\ \bibnamefont {Yang}},\
  and\ \bibinfo {author} {\bibfnamefont {C.}~\bibnamefont {Kim}},\ }\bibfield
  {title} {\bibinfo {title} {Efficient reactive brownian dynamics},\
  }\href@noop {} {\bibfield  {journal} {\bibinfo  {journal} {The Journal of
  chemical physics}\ }\textbf {\bibinfo {volume} {148}},\ \bibinfo {pages}
  {034103} (\bibinfo {year} {2018})}\BibitemShut {NoStop}%
\bibitem [{\citenamefont {Andrews}\ \emph {et~al.}(2010)\citenamefont
  {Andrews}, \citenamefont {Addy}, \citenamefont {Brent},\ and\ \citenamefont
  {Arkin}}]{andrews2010detailed}%
  \BibitemOpen
  \bibfield  {author} {\bibinfo {author} {\bibfnamefont {S.~S.}\ \bibnamefont
  {Andrews}}, \bibinfo {author} {\bibfnamefont {N.~J.}\ \bibnamefont {Addy}},
  \bibinfo {author} {\bibfnamefont {R.}~\bibnamefont {Brent}},\ and\ \bibinfo
  {author} {\bibfnamefont {A.~P.}\ \bibnamefont {Arkin}},\ }\bibfield  {title}
  {\bibinfo {title} {Detailed simulations of cell biology with smoldyn 2.1},\
  }\href@noop {} {\bibfield  {journal} {\bibinfo  {journal} {PLoS computational
  biology}\ }\textbf {\bibinfo {volume} {6}},\ \bibinfo {pages} {e1000705}
  (\bibinfo {year} {2010})}\BibitemShut {NoStop}%
\bibitem [{\citenamefont {van Zon}\ and\ \citenamefont
  {Ten~Wolde}(2005{\natexlab{a}})}]{van2005green}%
  \BibitemOpen
  \bibfield  {author} {\bibinfo {author} {\bibfnamefont {J.~S.}\ \bibnamefont
  {van Zon}}\ and\ \bibinfo {author} {\bibfnamefont {P.~R.}\ \bibnamefont
  {Ten~Wolde}},\ }\bibfield  {title} {\bibinfo {title} {Green’s-function
  reaction dynamics: a particle-based approach for simulating biochemical
  networks in time and space},\ }\href@noop {} {\bibfield  {journal} {\bibinfo
  {journal} {The Journal of chemical physics}\ }\textbf {\bibinfo {volume}
  {123}},\ \bibinfo {pages} {234910} (\bibinfo {year}
  {2005}{\natexlab{a}})}\BibitemShut {NoStop}%
\bibitem [{\citenamefont {van Zon}\ and\ \citenamefont
  {Ten~Wolde}(2005{\natexlab{b}})}]{van2005simulating}%
  \BibitemOpen
  \bibfield  {author} {\bibinfo {author} {\bibfnamefont {J.~S.}\ \bibnamefont
  {van Zon}}\ and\ \bibinfo {author} {\bibfnamefont {P.~R.}\ \bibnamefont
  {Ten~Wolde}},\ }\bibfield  {title} {\bibinfo {title} {Simulating biochemical
  networks at the particle level and in time and space: Green’s function
  reaction dynamics},\ }\href@noop {} {\bibfield  {journal} {\bibinfo
  {journal} {Physical review letters}\ }\textbf {\bibinfo {volume} {94}},\
  \bibinfo {pages} {128103} (\bibinfo {year} {2005}{\natexlab{b}})}\BibitemShut
  {NoStop}%
\bibitem [{\citenamefont {Sokolowski}\ \emph {et~al.}(2019)\citenamefont
  {Sokolowski}, \citenamefont {Paijmans}, \citenamefont {Bossen}, \citenamefont
  {Miedema}, \citenamefont {Wehrens}, \citenamefont {Becker}, \citenamefont
  {Kaizu}, \citenamefont {Takahashi}, \citenamefont {Dogterom},\ and\
  \citenamefont {ten Wolde}}]{sokolowski2019egfrd}%
  \BibitemOpen
  \bibfield  {author} {\bibinfo {author} {\bibfnamefont {T.~R.}\ \bibnamefont
  {Sokolowski}}, \bibinfo {author} {\bibfnamefont {J.}~\bibnamefont
  {Paijmans}}, \bibinfo {author} {\bibfnamefont {L.}~\bibnamefont {Bossen}},
  \bibinfo {author} {\bibfnamefont {T.}~\bibnamefont {Miedema}}, \bibinfo
  {author} {\bibfnamefont {M.}~\bibnamefont {Wehrens}}, \bibinfo {author}
  {\bibfnamefont {N.~B.}\ \bibnamefont {Becker}}, \bibinfo {author}
  {\bibfnamefont {K.}~\bibnamefont {Kaizu}}, \bibinfo {author} {\bibfnamefont
  {K.}~\bibnamefont {Takahashi}}, \bibinfo {author} {\bibfnamefont
  {M.}~\bibnamefont {Dogterom}},\ and\ \bibinfo {author} {\bibfnamefont
  {P.~R.}\ \bibnamefont {ten Wolde}},\ }\bibfield  {title} {\bibinfo {title}
  {egfrd in all dimensions},\ }\href@noop {} {\bibfield  {journal} {\bibinfo
  {journal} {The Journal of chemical physics}\ }\textbf {\bibinfo {volume}
  {150}},\ \bibinfo {pages} {054108} (\bibinfo {year} {2019})}\BibitemShut
  {NoStop}%
\bibitem [{\citenamefont {Opplestrup}\ \emph {et~al.}(2006)\citenamefont
  {Opplestrup}, \citenamefont {Bulatov}, \citenamefont {Gilmer}, \citenamefont
  {Kalos},\ and\ \citenamefont {Sadigh}}]{opplestrup2006first}%
  \BibitemOpen
  \bibfield  {author} {\bibinfo {author} {\bibfnamefont {T.}~\bibnamefont
  {Opplestrup}}, \bibinfo {author} {\bibfnamefont {V.~V.}\ \bibnamefont
  {Bulatov}}, \bibinfo {author} {\bibfnamefont {G.~H.}\ \bibnamefont {Gilmer}},
  \bibinfo {author} {\bibfnamefont {M.~H.}\ \bibnamefont {Kalos}},\ and\
  \bibinfo {author} {\bibfnamefont {B.}~\bibnamefont {Sadigh}},\ }\bibfield
  {title} {\bibinfo {title} {First-passage monte carlo algorithm: diffusion
  without all the hops},\ }\href@noop {} {\bibfield  {journal} {\bibinfo
  {journal} {Physical review letters}\ }\textbf {\bibinfo {volume} {97}},\
  \bibinfo {pages} {230602} (\bibinfo {year} {2006})}\BibitemShut {NoStop}%
\bibitem [{\citenamefont {Gillespie}\ \emph {et~al.}(2014)\citenamefont
  {Gillespie}, \citenamefont {Seitaridou},\ and\ \citenamefont
  {Gillespie}}]{gillespie2014small}%
  \BibitemOpen
  \bibfield  {author} {\bibinfo {author} {\bibfnamefont {D.~T.}\ \bibnamefont
  {Gillespie}}, \bibinfo {author} {\bibfnamefont {E.}~\bibnamefont
  {Seitaridou}},\ and\ \bibinfo {author} {\bibfnamefont {C.~A.}\ \bibnamefont
  {Gillespie}},\ }\bibfield  {title} {\bibinfo {title} {The small-voxel
  tracking algorithm for simulating chemical reactions among diffusing
  molecules},\ }\href@noop {} {\bibfield  {journal} {\bibinfo  {journal} {The
  Journal of chemical physics}\ }\textbf {\bibinfo {volume} {141}},\ \bibinfo
  {pages} {12B649\_1} (\bibinfo {year} {2014})}\BibitemShut {NoStop}%
\bibitem [{\citenamefont {Chew}\ \emph {et~al.}(2018)\citenamefont {Chew},
  \citenamefont {Kaizu}, \citenamefont {Watabe}, \citenamefont {Muniandy},
  \citenamefont {Takahashi},\ and\ \citenamefont {Arjunan}}]{chew2018reaction}%
  \BibitemOpen
  \bibfield  {author} {\bibinfo {author} {\bibfnamefont {W.-X.}\ \bibnamefont
  {Chew}}, \bibinfo {author} {\bibfnamefont {K.}~\bibnamefont {Kaizu}},
  \bibinfo {author} {\bibfnamefont {M.}~\bibnamefont {Watabe}}, \bibinfo
  {author} {\bibfnamefont {S.~V.}\ \bibnamefont {Muniandy}}, \bibinfo {author}
  {\bibfnamefont {K.}~\bibnamefont {Takahashi}},\ and\ \bibinfo {author}
  {\bibfnamefont {S.~N.}\ \bibnamefont {Arjunan}},\ }\bibfield  {title}
  {\bibinfo {title} {Reaction-diffusion kinetics on lattice at the microscopic
  scale},\ }\href@noop {} {\bibfield  {journal} {\bibinfo  {journal} {Physical
  Review E}\ }\textbf {\bibinfo {volume} {98}},\ \bibinfo {pages} {032418}
  (\bibinfo {year} {2018})}\BibitemShut {NoStop}%
\bibitem [{\citenamefont {Von~Smoluchowski}(1917)}]{von1917mathematical}%
  \BibitemOpen
  \bibfield  {author} {\bibinfo {author} {\bibfnamefont {M.}~\bibnamefont
  {Von~Smoluchowski}},\ }\bibfield  {title} {\bibinfo {title} {Mathematical
  theory of the kinetics of the coagulation of colloidal solutions},\
  }\href@noop {} {\bibfield  {journal} {\bibinfo  {journal} {Z. Phys. Chem}\
  }\textbf {\bibinfo {volume} {92}},\ \bibinfo {pages} {129} (\bibinfo {year}
  {1917})}\BibitemShut {NoStop}%
\bibitem [{\citenamefont {Collins}\ and\ \citenamefont
  {Kimball}(1949)}]{collins1949diffusion}%
  \BibitemOpen
  \bibfield  {author} {\bibinfo {author} {\bibfnamefont {F.~C.}\ \bibnamefont
  {Collins}}\ and\ \bibinfo {author} {\bibfnamefont {G.~E.}\ \bibnamefont
  {Kimball}},\ }\bibfield  {title} {\bibinfo {title} {Diffusion-controlled
  reaction rates},\ }\href@noop {} {\bibfield  {journal} {\bibinfo  {journal}
  {Journal of colloid science}\ }\textbf {\bibinfo {volume} {4}},\ \bibinfo
  {pages} {425} (\bibinfo {year} {1949})}\BibitemShut {NoStop}%
\bibitem [{\citenamefont {Naqvi}\ \emph {et~al.}(1982)\citenamefont {Naqvi},
  \citenamefont {Waldenstr{\o}m},\ and\ \citenamefont
  {Mork}}]{naqvi1982kinetics}%
  \BibitemOpen
  \bibfield  {author} {\bibinfo {author} {\bibfnamefont {K.~R.}\ \bibnamefont
  {Naqvi}}, \bibinfo {author} {\bibfnamefont {S.}~\bibnamefont
  {Waldenstr{\o}m}},\ and\ \bibinfo {author} {\bibfnamefont {K.}~\bibnamefont
  {Mork}},\ }\bibfield  {title} {\bibinfo {title} {Kinetics of
  diffusion-mediated bimolecular reactions. a new theoretical framework},\
  }\href@noop {} {\bibfield  {journal} {\bibinfo  {journal} {The Journal of
  Physical Chemistry}\ }\textbf {\bibinfo {volume} {86}},\ \bibinfo {pages}
  {4750} (\bibinfo {year} {1982})}\BibitemShut {NoStop}%
\bibitem [{\citenamefont {Noyes}(1956)}]{noyes1956models}%
  \BibitemOpen
  \bibfield  {author} {\bibinfo {author} {\bibfnamefont {R.}~\bibnamefont
  {Noyes}},\ }\bibfield  {title} {\bibinfo {title} {Models relating molecular
  reactivity and diffusion in liquids},\ }\href@noop {} {\bibfield  {journal}
  {\bibinfo  {journal} {Journal of the American Chemical Society}\ }\textbf
  {\bibinfo {volume} {78}},\ \bibinfo {pages} {5486} (\bibinfo {year}
  {1956})}\BibitemShut {NoStop}%
\bibitem [{\citenamefont {Noyes}(1961)}]{noyes1961prog}%
  \BibitemOpen
  \bibfield  {author} {\bibinfo {author} {\bibfnamefont {R.}~\bibnamefont
  {Noyes}},\ }\bibfield  {title} {\bibinfo {title} {Effects of diffusion rates
  on chemical kinetics},\ }in\ \href@noop {} {\emph {\bibinfo {booktitle}
  {Progress in Reaction Kinetics}}},\ Vol.~\bibinfo {volume} {1},\ \bibinfo
  {editor} {edited by\ \bibinfo {editor} {\bibfnamefont {S.}~\bibnamefont
  {Benson}}}\ (\bibinfo {year} {1961})\ pp.\ \bibinfo {pages}
  {129--160}\BibitemShut {NoStop}%
\bibitem [{\citenamefont {Zhang}\ \emph {et~al.}(2005)\citenamefont {Zhang},
  \citenamefont {Rohlfs},\ and\ \citenamefont
  {Schwartz}}]{zhang2005implementation}%
  \BibitemOpen
  \bibfield  {author} {\bibinfo {author} {\bibfnamefont {T.}~\bibnamefont
  {Zhang}}, \bibinfo {author} {\bibfnamefont {R.}~\bibnamefont {Rohlfs}},\ and\
  \bibinfo {author} {\bibfnamefont {R.}~\bibnamefont {Schwartz}},\ }\bibfield
  {title} {\bibinfo {title} {Implementation of a discrete event simulator for
  biological self-assembly systems},\ }in\ \href@noop {} {\emph {\bibinfo
  {booktitle} {Proceedings of the 37th conference on Winter simulation}}}\
  (\bibinfo {organization} {Winter Simulation Conference},\ \bibinfo {year}
  {2005})\ pp.\ \bibinfo {pages} {2223--2231}\BibitemShut {NoStop}%
\bibitem [{\citenamefont {Anderson}(2007)}]{anderson2007modified}%
  \BibitemOpen
  \bibfield  {author} {\bibinfo {author} {\bibfnamefont {D.~F.}\ \bibnamefont
  {Anderson}},\ }\bibfield  {title} {\bibinfo {title} {A modified next reaction
  method for simulating chemical systems with time dependent propensities and
  delays},\ }\href@noop {} {\bibfield  {journal} {\bibinfo  {journal} {The
  Journal of chemical physics}\ }\textbf {\bibinfo {volume} {127}},\ \bibinfo
  {pages} {214107} (\bibinfo {year} {2007})}\BibitemShut {NoStop}%
\bibitem [{\citenamefont {Mathai}\ and\ \citenamefont
  {Provost}(1992)}]{mathai1992quadratic}%
  \BibitemOpen
  \bibfield  {author} {\bibinfo {author} {\bibfnamefont {A.~M.}\ \bibnamefont
  {Mathai}}\ and\ \bibinfo {author} {\bibfnamefont {S.~B.}\ \bibnamefont
  {Provost}},\ }\href@noop {} {\emph {\bibinfo {title} {Quadratic forms in
  random variables: theory and applications}}}\ (\bibinfo  {publisher}
  {Dekker},\ \bibinfo {year} {1992})\BibitemShut {NoStop}%
\bibitem [{\citenamefont {Thomas}(2020)}]{dessaCS_github}%
  \BibitemOpen
  \bibfield  {author} {\bibinfo {author} {\bibfnamefont {M.}~\bibnamefont
  {Thomas}},\ }\href {https://doi.org/10.5281/zenodo.3964563} {\bibinfo {title}
  {{DESSA-CS}}} (\bibinfo {year} {2020})\BibitemShut {NoStop}%
\bibitem [{\citenamefont {Carslaw}\ and\ \citenamefont
  {Jaeger}(1959)}]{carslaw1959conduction}%
  \BibitemOpen
  \bibfield  {author} {\bibinfo {author} {\bibfnamefont {H.~S.}\ \bibnamefont
  {Carslaw}}\ and\ \bibinfo {author} {\bibfnamefont {J.~C.}\ \bibnamefont
  {Jaeger}},\ }\bibfield  {title} {\bibinfo {title} {Conduction of heat in
  solids},\ }\href@noop {} {\bibfield  {journal} {\bibinfo  {journal} {Oxford:
  Clarendon Press, 1959, 2nd ed.}\ } (\bibinfo {year} {1959})}\BibitemShut
  {NoStop}%
\bibitem [{\citenamefont {Van~Beijeren}\ \emph {et~al.}(2001)\citenamefont
  {Van~Beijeren}, \citenamefont {Dong},\ and\ \citenamefont
  {Bocquet}}]{van2001diffusion}%
  \BibitemOpen
  \bibfield  {author} {\bibinfo {author} {\bibfnamefont {H.}~\bibnamefont
  {Van~Beijeren}}, \bibinfo {author} {\bibfnamefont {W.}~\bibnamefont {Dong}},\
  and\ \bibinfo {author} {\bibfnamefont {L.}~\bibnamefont {Bocquet}},\
  }\bibfield  {title} {\bibinfo {title} {Diffusion-controlled reactions: A
  revisit of noyes’ theory},\ }\href@noop {} {\bibfield  {journal} {\bibinfo
  {journal} {The Journal of Chemical Physics}\ }\textbf {\bibinfo {volume}
  {114}},\ \bibinfo {pages} {6265} (\bibinfo {year} {2001})}\BibitemShut
  {NoStop}%
\bibitem [{\citenamefont {Kaizu}\ \emph {et~al.}()\citenamefont {Kaizu},
  \citenamefont {Nishida}, \citenamefont {Sakamoto}, \citenamefont {Kato},
  \citenamefont {Niina}, \citenamefont {Nishida}, \citenamefont {Koizumi},
  \citenamefont {Aota}, ,\ and\ \citenamefont {Takahashi}}]{ecell4}%
  \BibitemOpen
  \bibfield  {author} {\bibinfo {author} {\bibfnamefont {K.}~\bibnamefont
  {Kaizu}}, \bibinfo {author} {\bibfnamefont {K.}~\bibnamefont {Nishida}},
  \bibinfo {author} {\bibfnamefont {Y.}~\bibnamefont {Sakamoto}}, \bibinfo
  {author} {\bibfnamefont {S.}~\bibnamefont {Kato}}, \bibinfo {author}
  {\bibfnamefont {T.}~\bibnamefont {Niina}}, \bibinfo {author} {\bibfnamefont
  {N.}~\bibnamefont {Nishida}}, \bibinfo {author} {\bibfnamefont
  {M.}~\bibnamefont {Koizumi}}, \bibinfo {author} {\bibfnamefont
  {N.}~\bibnamefont {Aota}}, ,\ and\ \bibinfo {author} {\bibfnamefont
  {K.}~\bibnamefont {Takahashi}},\ }\href
  {https://doi.org/10.5281/zenodo.1119017} {\bibinfo {title} {{E-Cell version
  4}}}\BibitemShut {NoStop}%
\end{thebibliography}%

\end{document}